\documentclass[usenatbib]{mnras}
\usepackage{graphics}
\usepackage{lipsum}
\usepackage{amsmath}
\usepackage{breqn}
\usepackage{amssymb}
\usepackage{hyperref}
\hypersetup{draft} 
\usepackage{subfigure}
\usepackage{multicol}
\usepackage{caption}
\usepackage{multirow}
\usepackage{ulem}
\usepackage{rotating}
\usepackage{pdflscape}

\usepackage{color}

\title{HOD Modelling of High Redshift Galaxies Using the \texttt{BlueTides} Simulation}
 
\author[Bhowmick et al.]{
Aklant K. Bhowmick$^{1}$,
Duncan Campbell$^{1}$,
Tiziana DiMatteo$^{1}$ and Yu Feng$^{2}$ \\
$^{1}$McWilliams Center for Cosmology, Dept. of Physics, Carnegie Mellon University,
Pittsburgh PA 15213, USA\\
$^{2}$Berkeley Center for Cosmological Physics, University of California at Berkeley,
Berkeley CA 94720, USA
}

\date{Accepted XXX. Received YYY; in original form ZZZ}

\begin{document}
\maketitle
\begin{abstract}
We construct halo occupation distribution (HOD) models of high redshift ($z \gtrsim 7.5$) galaxies with $M_{*}>10^8~M_{\odot}/h$ using the \texttt{BlueTides} hydrodynamic simulation suite, with a particular emphasis on modelling the small scale / 1-halo clustering ($0.01\lesssim r \lesssim 1~ h^{-1}\rm{Mpc}$). Similar to low redshift studies, we find that the central and satellite mean HODs ($\left<N_{\mathrm{cen}}\right>$ and $\left<N_{\mathrm{sat}}\right>$) can be modeled by a smoothed step function and a power law respectively. The number density of satellite galaxies is however significantly suppressed compared to low redshift (satellite fractions drop from $\sim 50 \%$ at $z=0$ to $\lesssim 10 \%$ at $z=7.5$). The mean number of satellites, $\left<N_{\mathrm{sat}}\right> < 1$ for halo masses below $3 \times 10^{11} M_{\odot}/h$ (a rare halo at these redshifts). For the radial number density profiles, satellites with $10^8 \lesssim M^* \lesssim 10^{9} M_{\odot}/h$ in halos with $M_H \gtrsim 3 \times10^{11} M_{\odot}/h$ are consistent with NFW (with concentrations $c_{\mathrm{sat}} \sim 10-40$). 
Within halos of mass $M_H\lesssim 3 \times 10^{11} M_{\odot}/h$ satellites exhibit a power law profile with slope -3. Because these halos dominate the small scale clustering, the resulting 1-halo term is steeper than predicted using standard NFW profiles. Using this power-law profile for satellites, we can successfully reproduce the small-scale clustering exhibited by \texttt{BlueTides} galaxies using HOD modelling. We predict the highest probability of detecting satellites at $z>7.5$ is around centrals of $M^*\sim 3 \times 10^{10} M_{\odot}/h$ (with $M^{*}\gtrsim$ a few $10^{7} M_{\odot}/h$ ). This should be achievable with the James Webb Space Telescope (JWST).    
\end{abstract}


\section{Introduction}
\label{introduction}
Significant progress has been made in extending surveys of high redshift galaxies, opening up a new frontier for probing cosmology and galaxy formation physics. The first spectroscopic confirmations of high redshift Lyman Break Galaxies (LBGs) were at $z\sim 3-4$ using the W.M. Keck Telescope: Low Resolution Imaging Spectrograph (LRIS) \citep{1996ApJ...462L..17S}, and the Hubble Deep Field (HDF) \citep{1996MNRAS.283.1388M,1997AJ....113....1S}. The  Hubble Space Telescope- Advanced Camera for Surveys (HST-ACS) enabled the detection of galaxies up to $z\sim 6$ \citep{2003MNRAS.342..439S, 2004ApJ...611L...1B,2004ApJ...600L..99D}.  The installation of the Wide Field Camera 3 and near IR camera on the Hubble Space Telescope (HST-WFC3/ IR) lead to the identification of 200-500 galaxies at $z\sim 7-8$ \citep{2010MNRAS.403..938W,2011ApJ...737...90B,2012ApJ...756..164F,2012A&A...547A..51G,2012ApJ...759..135O,2012ApJ...761..177Y,2013MNRAS.429..150L,2013MNRAS.432.2696M,2013ApJ...768..196S,2014ApJ...786...57S}; HST-WFC3 has since been used to detect a handful of exceptionally bright galaxies at $z\gtrsim 9$ galaxies \citep{2014ApJ...786..108O}, including GN-z11 \citep{2016ApJ...819..129O}, the highest spectroscopically confirmed redshift galaxy observed to date. Additionally, deep ground based surveys from the Canada–France–Hawaii Telescope (CFHT), Subaru Suprime Cam (SSC), VLT and VISTA  have greatly contributed in complementing the wavelength coverage available from HST, extending from 3500 to 23000 \AA, enabling better identifications of galaxies from $z=5-10$ and making the samples less prone to contamination from low-redshift interlopers. 

High redshift catalogs have now become large enough to perform clustering measurements. Data from the Subaru Deep Survey, CFHT Legacy Survey, and Large Binocular Telescope (LBT) Bootes Field survey have been used to make clustering measurements at $z\sim 3-5$ \citep{2001ApJ...558L..83O,2004ApJ...611..660O,2004ApJ...611..685O,2009A&A...498..725H,2013ApJ...774...28B}. The combined compilation of the HST-WFC3 and the Subaru Hyper Suprime Cam (HSC) surveys has been used to measure clustering of LBGs at $z\sim 4-6$ \citep[hereafter Har16, Har17 ]{2016ApJ...821..123H,2017arXiv170406535H}. \cite[hereafter Hat17]{2017arXiv170203309H} uses bright LBG samples from wider area surveys such as Subaru XMM-Newton Deep Survey (SXDS) and the ultraVISTA survey to measure LBG clustering at $z\sim6$. Other high redshift clustering measurements at $z\sim 4-6$ include \cite{0004-637X-637-2-631}, \cite{1538-4357-648-1-L5}. While the limits of clustering analyses are continuely being pushed, the current observational limit is at $z\sim 7$, with measurements performed using the HST-WFC3 fields in \cite{2014ApJ...793...17B} and HST-WFC3+Subaru HSC fields in \cite{2016ApJ...821..123H}; but this limit will be pushed to whole new frontiers with upcoming missions such as the WFIRST and JWST \citep[and references therein]{2015JPhCS.610a2007G,2006SSRv..123..485G} which will reach unprecedented depths as well as sky coverage, revolutionizing the field of high redshift galaxy studies.   

Halo Occupation Distribution (HOD) modelling of galaxy clustering observations serves as a powerful tool to constrain the galaxy-dark matter halo connection \citep{2002ApJ...575..587B,2004ApJ...609...35K,2005ApJ...633..791Z}. Particularly in the high redshift regime, other probes, such as weak lensing, 
\citep{2005MNRAS.361.1287M,2007JCAP...06..024M,2015MNRAS.452.2225G} are difficult to implement due to the lack of background galaxies at higher redshifts; and subhalo abundance matching (SHAM) \citep{2004MNRAS.353..189V,2006ApJ...647..201C,2006ApJ...652...71W,2006ApJ...643...14S,2016MNRAS.460.3100C} traditionally assumes low or no scatter between the (sub-)halo mass and any galaxy property such as stellar mass or luminosity (as we will see the scatter may be significantly higher at higher redshifts). Therefore, in the high redshift regime, traditional clustering analyses may be the most robust tool to probe the galaxy-halo connection.

HOD modeling, or the closely realted conditional luminosity function (CLF) formalism, assume a probabilistic connection between host halos and the galaxies they occupy. It can be easily incorporated into the halo model framework to provide an analytical prediction for galaxy clustering \citep{2000MNRAS.318..203S,2002PhR...372....1C,2003MNRAS.339.1057Y,2012ApJ...761..127M}.  \cite{2002ApJ...575..587B} showed that the HOD model parameters and the cosmological parameters have non-degenerate effects on the galaxy clustering, implying that galaxy clustering measurements can (atleast in principle) simultaneously constrain the cosmology as well as HOD \citep{2005ApJ...625..613A,2012ApJ...745...16T,2013MNRAS.430..767C,2013MNRAS.430..725V,2014ApJ...783..118R}. Furthermore, given a cosmology, HODs of galaxies of different types (color, stellar mass, luminosity, star formation rate etc.) is completely determined by the physics of galaxy formation, implying that HOD modelling can be used to test galaxy formation theories \citep{2002ApJ...575..587B,2007ApJ...667..760Z,2013MNRAS.429.3604B,2016arXiv160800370M}. HOD modelling has therefore been extensively applied to low redshifts $z\sim 0-2$ \citep[e.g.][and references therein]{0004-637X-831-1-3,2015MNRAS.449..901M,2015MNRAS.449.1352C}.      

Recently, HOD modelling has been applied to clustering observations in the high redshift regime, $z\sim 4-7$ \citep{2016ApJ...821..123H,2017arXiv170203309H,2017arXiv170406535H}. However, the parametrizations used in most HOD modelling are motivated from simulations at low redshifts \citep{2002ApJ...575..587B,2004ApJ...609...35K}. Commonly used parametrizations have 5-6 free parameters. Furthermore, the statistical power of clustering measurements in this regime are still not enough to simulataneously constrain such a large number of parameters, which makes it necessary to adopt additional constraints (see for example Eqs. (54) and (55) and additional parameters with fixed values in \cite{2016ApJ...821..123H}). We must also note that constraining galaxy formation directly with simulations is still a very daunting task. In other words, running a large number of simulations with different galaxy formation physics to see which one best describes the observations, is not computationally feasible at this point. This is also true in the high redshift regime where galaxies become increasingly rare and compact, requiring higher volume and resolution simulations for analysing their statistics. Therefore, even with the next generation of faster supercomputers, high redshift galaxies will continue to be a regime where analytical models such as the HOD model will serve as an indispensable tool to constrain galaxy formation physics. 

Recent work using \texttt{BlueTides} \citep{2017arXiv170702312B} and semi-analytic (SA) modelling \citep{2017MNRAS.472.1995P} suggests that on large scales (in the two-halo regime) standard HOD modelling assumptions work well at these high redshifts, where the inferred halo mass estimates of galaxies \citep{2016ApJ...821..123H, 2017arXiv170406535H} are consistent with \texttt{BlueTides} predictions (despite the exclusion of the non-linear bias effect). However, on small scales (in the one-halo regime) we found enhanced clustering compared to standard HOD assumptions which assume an NFW profile. Additionally, some of the very first attempts \citep{2017arXiv170203309H,2017arXiv170406535H} of fitting the one-halo clustering measurements to HOD model predictions led to inferred satellite abundances significantly lower than seen in simulations. These findings suggest that as we probe higher redshifts with clustering analyses, it is important to carefully validate assumptions in HOD models, and to identify suitable modifications where required.

Testing the basic assumptions in HOD modelling requires information about how halos connect to galaxies. This needs to come from galaxy formation physics \citep[and references therein]{1999ApJ...513..142M,2014MNRAS.445.2545D,0004-637X-836-2-204}. Galaxy formation physics is employed and coupled with dark matter either by post processing of dark matter only simulations with Semi-Analytical (SA) approach \citep{1991ApJ...379...52W,1993MNRAS.264..201K,1994MNRAS.271..781C,1998ApJ...505...37A,1999MNRAS.310.1087S}, or by incorporation of gas dynamics along with dark matter in the form of hydrodynamic simulations \citep{1992ApJ...399L.113C,1992ApJ...399L.109K,1994ApJ...422...11E,1999ApJ...521L..99P,2001ApJ...550L.129W,2001ApJ...558..520Y,2012ApJ...745L..29D,Nelson201512,2015MNRAS.450.1349K, 2016MNRAS.455.2778F}. Some studies which have used N-body, N-body+SA or N-body+hydrodynamic simulations to build/constrain low to medium redshift ($z\sim0-5$) HOD models and its key components include \cite{2000MNRAS.311..793B}, \cite{2001ApJ...550L.129W}, \cite{2001ApJ...558..520Y}, \cite{2003ApJ...593....1B}, \cite{2004ApJ...609...35K}, \cite{2006ApJ...647..201C}, \cite{2012MNRAS.419.2657C}; these studies investigate in detail various properties of HODs and other relevant galaxy properties, which motivated a set of standard parametrizations and assumptions and are now widely used in clustering analysis using HOD modelling.  

In this work, we investigate the galaxy population of \texttt{BlueTides}, and build HOD model components for $z\gtrsim7.5$ galaxies by using \texttt{BlueTides}, a recent large volume, high resolution cosmological hydrodynamic simulation \citep{2016MNRAS.455.2778F}. We shall put particular emphasis on the satellite galaxies and the modelling of one-halo clustering. As we shall see in the details of section \ref{simulation}, \texttt{BlueTides} uniquely large simulated volume provides us with sufficient number of  high redshift ($z\gtrsim 7.5$) galaxies to perform statistical studies, making it a good testing ground for validating HOD models in this high redshift domain.  

\section{Methods}
\subsection{BlueTides Simulation}
\label{simulation}

\texttt{BlueTides} is a high resolution cosmological hydrodynamic simulation 
that has been run to $z\sim 7.5$ on the Bluewaters supercomputer. 
\texttt{BlueTides} makes use of  \texttt{MP-GADGET} which employs the Pressure entropy formulation of Smooth Particle Hydrodynamics. It has a simulation box size of $400^{3}$ $(Mpc/h)^{3}$, (almost 200 times the volume of \texttt{ILLUSTRIS} \citep{2014MNRAS.445..175G} as well as \texttt{EAGLE} \citep{2015MNRAS.446..521S}); with $2\times 7048^{3}$ particles, the resolution is comparable to that of \texttt{ILLUSTRIS}, making it the largest high resolution hydrodynamic simulation for the given volume run to date. The large volume of the simulation allows for the formation of rare massive halos hosting bright galaxies, making it ideally suited for the high-redshift regime where only the brightest galaxies are detectable in the current surveys. The cosmological parameters adopted are from the nine-year Wilkinson Microwave Anisotropy Probe (WMAP) \citep{2013ApJS..208...19H} ($\Omega_0=0.2814$, $\Omega_\lambda=0.7186$, $\Omega_b=0.0464$ $\sigma_8=0.82$, $h=0.697$, $n_s=0.971$). 
For more details on \texttt{BlueTides}, interested reader should refer to \cite{2016MNRAS.455.2778F}. 

The various sub-grid physics models that have been employed in \texttt{BlueTides} are:
\begin{itemize}
\item{Multiphase star formation model \citep{2003MNRAS.339..289S,2013MNRAS.436.3031V}}
\item{Molecular hydrogen formation \citep{2011ApJ...729...36K}}
\item{Gas cooling via radiative transfer \citep{1996ApJS..105...19K} and metal cooling \citep{2014MNRAS.444.1518V}}
\item{SNII feedback \citep{Nelson201512}}
\item{Black hole growth and AGN feedback \citep{2005MNRAS.361..776S,2005Natur.433..604D}}
\item{``Patchy" reionization \citep{2013ApJ...776...81B}}
\end{itemize}

\texttt{BlueTides} has revealed significant insights into the properties and evolution of high redshift galaxies. 
The UV luminosity functions at $z=8,9,10$ \citep{2015ApJ...808L..17F, 2016MNRAS.455.2778F,
2016MNRAS.463.3520W} are consistent with the observational constraints \citep{2015ApJ...803...34B}. \texttt{BlueTides} sucessfully predicted the properties of GN-z11, the farthest galaxy observed till date \citep{2016MNRAS.461L..51W}. Clustering properties of \texttt{BlueTides} galaxies at $z\sim8,9,10$ are also consistent with currently available observations \citep{2017arXiv170702312B}. The large volume provided by \texttt{BlueTides} has allowed the study of the rare earliest supermassive black holes/first quasars and the role of tidal field in the black hole growth in the early universe \citep{2017MNRAS.467.4243D}. Dark matter only realization of \texttt{BlueTides} (MassTracer simulations) has been used to study the descendants of the first quasars to the present day \citep{2017arXiv170803373T}. Photometric properties of high redshift galaxies predicted by \texttt{BlueTides} and their dependence on the choice of stellar population synthesis modeling as well as dust modeling have been extensively studied in \cite{2016MNRAS.458L...6W, 2016MNRAS.460.3170W, 2018MNRAS.473.5363W}.
By extending the simulation to $z=7.5$ we have also been able to confront predictions from \texttt{BlueTides} (Tenneti et. al. 2018 in prep, Ni et. al. 2018 in prep) for
the recently discovered highest redshift quasar \citep{2018Natur.553..473B}.

\subsubsection{Dark Matter Halos and Sub-Halos}
We  two distinct halo catalogs for our snapshots: 1) Friends of friends (FOF) halos \citep{1985ApJ...292..371D}, generate on-the-fly, defined by linking together pairs of particles within \texttt{LINKING_LENGTH=0.2} times the mean particle spacing  2) Spherical overdense (SO) halos centered at the density peaks (in phase space) and with average matter density within the halo boundary being equal to 200 times the mean density of the universe; this is done using \texttt{ROCKSTAR} halo finder \citep{2013ApJ...762..109B} which allows
us to identify subtructure as well. 
For this reason we make use SO halos
and subhalos in this work (c.f. \S 2.1.2)

To validate our halo catalogue we check on the resulting halo mass functions, that they agree well with analytical fitting functions \citep{2013ApJ...762..109B} used in recent works which perform HOD modelling at high redshifts. The comparison is discussed in Appendix \ref{mass_function}. 

\subsubsection{Identifying Central and Satellite Galaxies}
\begin{table}
\begin{center}
\begin{tabular}{|c|c|c|c|}
$z$ & $\log_{10}M^*_{th}h^{-1}M_{\odot}$ & $N_{\mathrm{central}}$ & $N_{\mathrm{satellites}}$ \\

\hline
 $7.5$& $8.0$&  $171405$  & $13711$ \\
 $7.5$& $8.5$ & 	$45923$ &  $2524$  \\
 $7.5$& $9.0$ & 	$10227$ &  $279$   \\
 $7.5$& $9.5$ & $1671$ &  $22$\\ 
  $7.5$& $10.0$ & $186$ &  $1$\\ 
   $7.5$& $10.5$ & $8$ &  $0$\\ 
 \hline
 $8$& $8.0$&  $94596$  & $6782$ \\
 $8$& $8.5$ & 	$22115$ &  $1018$  \\
 $8$& $9.0$ & 	$4101$ &  $94$   \\
 $8$& $9.5$ & $548$ &  $9$\\ 
  $8$& $10.0$ & $46$ &  $0$\\ 
   $8$& $10.5$ & $4$ &  $0$\\ 
  \hline
 $9$& $8.0$&  $27360$  & $1394$ \\
 $9$& $8.5$ & 	$4994$ &  $173$  \\
 $9$& $9.0$ & 	$702$ &  $10$   \\
 $9$& $9.5$ & $72$ &  $1$\\ 
  $9$& $10.0$ & $6$ &  $0$\\ 
 \hline
 $10$& $8.0$&  $6803$  & $285$ \\
 $10$& $8.5$ & 	$1005$ &  $23$  \\
 $10$& $9.0$ & 	$117$ &  $2$   \\
 $10$& $9.5$ & $5$ &  $0$\\ 
\end{tabular}
\end{center}
\caption{Stellar mass thresholds and the number of central and satellite galaxies identified in \texttt{BlueTides} at redshifts 7.5,8,9,10.}
\label{galaxy_samples}
\end{table}

Star particles are uniquely assigned to the closest density peak (halo or subhalo) identified by \texttt{ROCKSTAR}. Within each host halo, the most massive galaxy is defined to be the central galaxy, while all others are defined to be satellite galaxies. Galaxy pairs which are found to be closer than 8 ckpcs ($\sim$ typical half-mass radius of these galaxies) are considered to be 'merged', and we combine these into a single object. Galaxy centers are defined to be the center of masses of the stellar distributions of galaxies. Figure \ref{galaxy_snaps} shows some examples of the stellar mass distributions of central (red histograms) and satellite (blue histograms) galaxies within host haloes. The number of central and satellite galaxies identified in \texttt{BlueTides} for various stellar mass thresholds is listed in Table \ref{galaxy_samples}. 

\subsection{MassiveBlack II Simulation}
\texttt{MassiveBlack II} (MBII) \citep[for details]{2015MNRAS.450.1349K} is  a high resolution cosmological hydrodynamic simulation with a boxsize of $100 Mpc/h$ and $2 \times 1792^3$ particles. The simulation was run to $z\sim 0$  with cosmological parameters consistent with WMAP7 \citep{2011ApJS..192...18K} ($\Omega_0=0.275$, $\Omega_\lambda=0.725$, $\Omega_b=0.046$ $\sigma_8=0.816$, $h=0.701$, $n_s=0.968$). MBII was
run with an earlier version of the
MP-GADGET code and besides hydrodynamics and gravity it includes a multiphase star formation model with a (somewhat simpler) supernova feedback and modelling of black hole growth and associated AGN feedback
MBII has been used and validated in a variety  studies a wide that range of galaxy and black hole properties such as galaxy and black hole luminosity functions, two-point clustering, HODs \citep{2015MNRAS.450.1349K}, emission lines \citep{2015MNRAS.454..269P}, galaxy shapes and intrinsic allignments \citep{2015MNRAS.453..469T,2016MNRAS.462.2668T} and radial acceleration of disk galaxies \citep{2016MNRAS.462.2668T}. 
In this work, we use MBII to compare HOD predictions for galaxy populations, particularly the abundance of satellite galaxies between low ($z\sim0$) and high redshifts ($z\gtrsim7.5$). However, note that the comparison should be intended mostly as illustrative and not meant to be strictly quantitative due to differences in cosmology and details in subgrid modeling between the two simulations.



 \section{Central and Satellite galaxy populations in BlueTides}
\begin{figure*}

\begin{tabular}{cccc}
    \includegraphics[width=4cm, height=4cm]{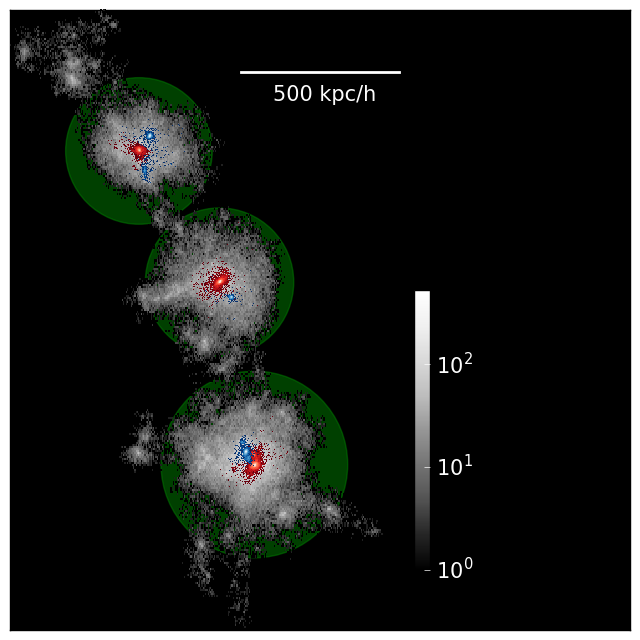}&
    \includegraphics[width=4cm, height=4cm]{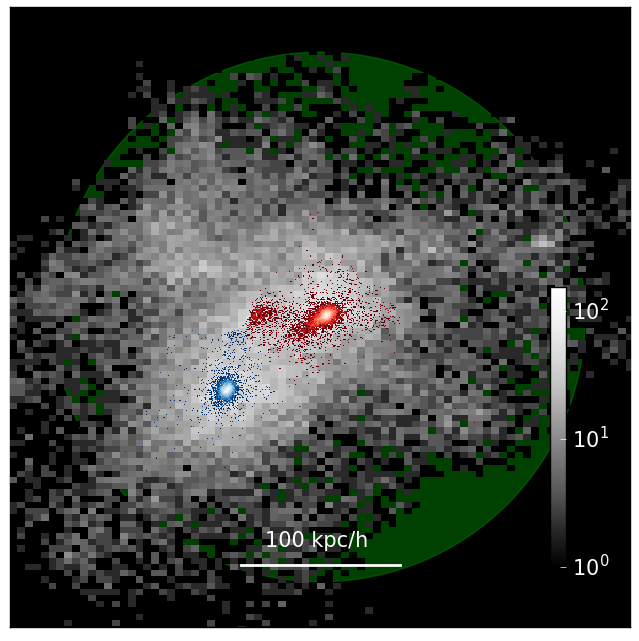}&
    \includegraphics[width=4cm, height=4cm]{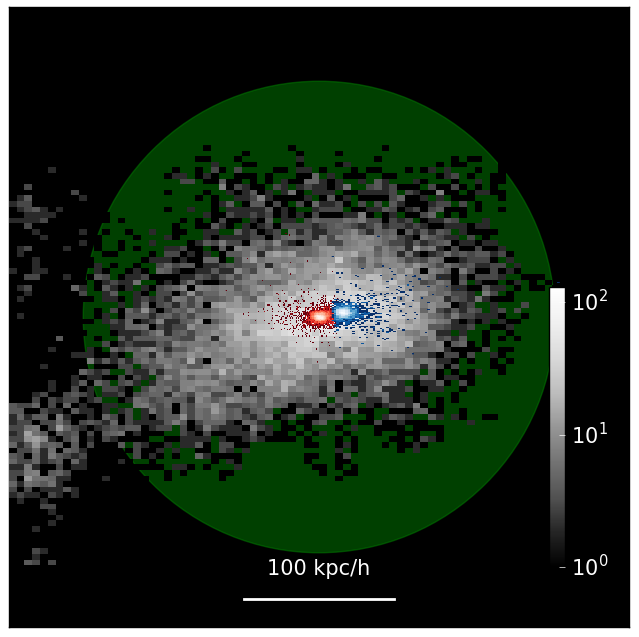}&
    \includegraphics[width=4cm, height=4cm]{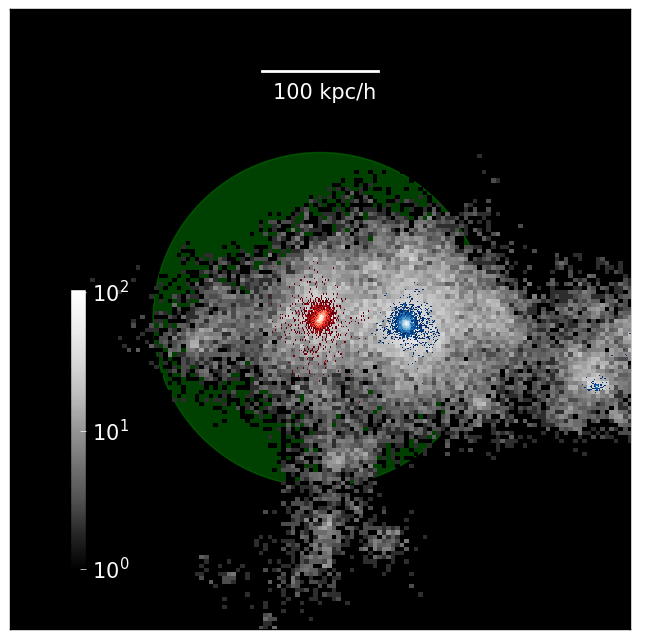}\\
\end{tabular}
\caption{Illustrative examples of stellar distributions of central (red histograms) and satellite (blue histograms) galaxies ($M^*>10^9 M_{\odot}/h$) identified in \texttt{BlueTides}. The grey histograms show the dark matter distributions of the parent FOFGroup. The grey circles denote the virial radii of the corresponding host SO halos identified by \texttt{ROCKSTAR}. The color bars denote the number of particles in a bin (see legend).}
\label{galaxy_snaps}. 

\begin{tabular}{cc}
    \includegraphics[width=9cm, height=6.3cm]{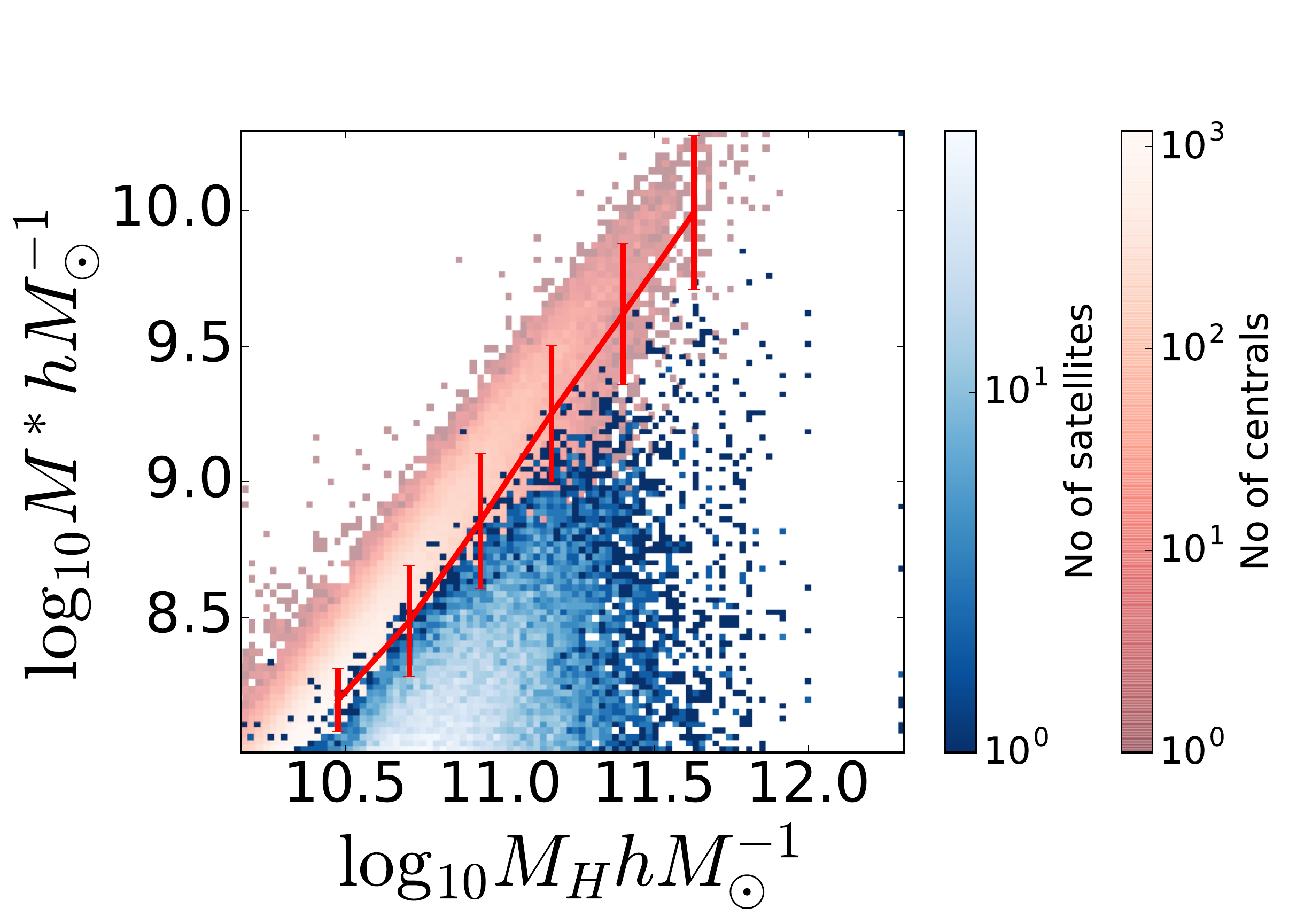}&

    \includegraphics[width=7cm, height=5.5cm]{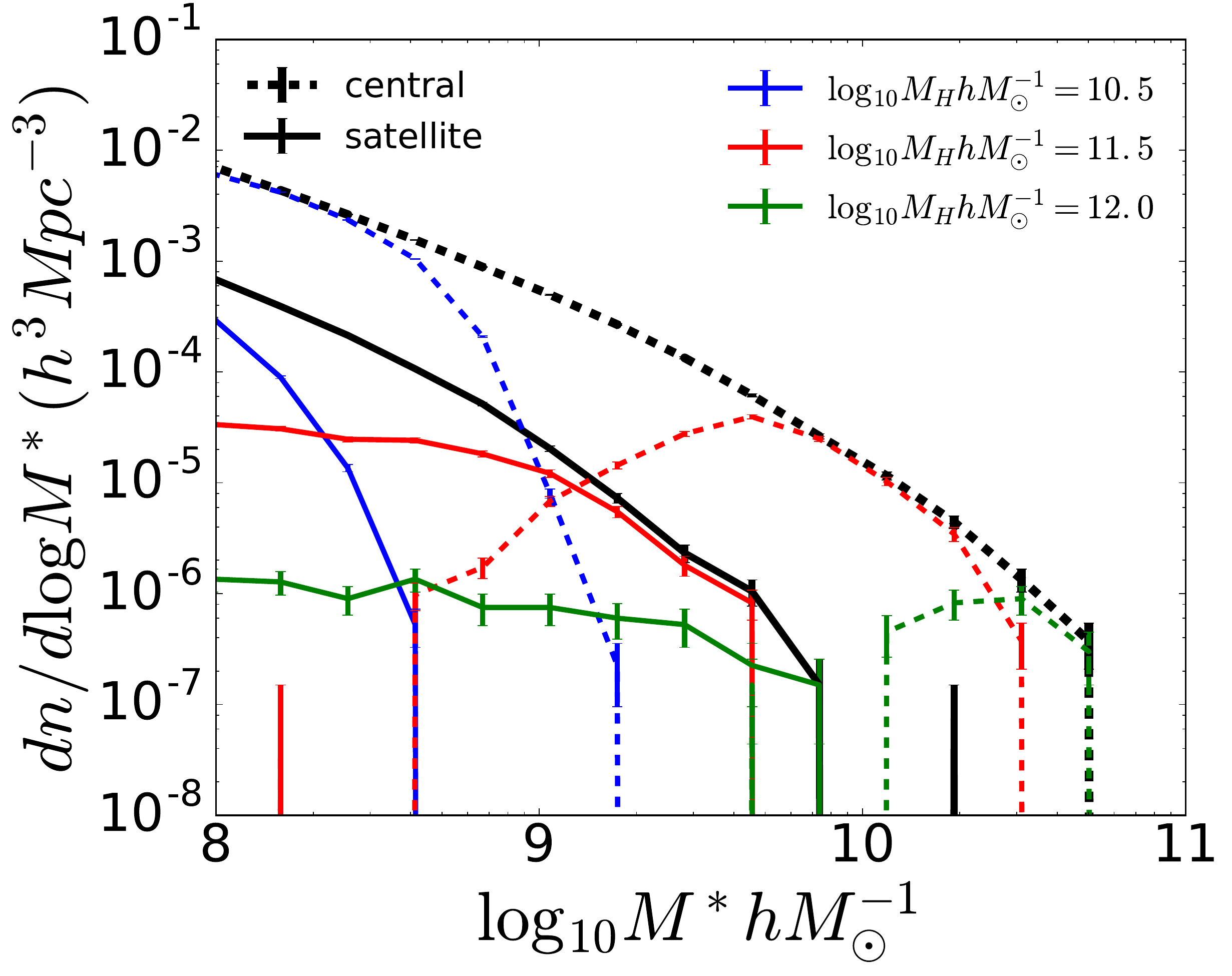}\\
\end{tabular}
\caption{Stellar to halo mass relations of central (blue histogram) and satellite (red histogram) galaxies predicted by \texttt{BlueTides} at $z=7.5$ The red line shows the mean relation for central galaxies with $1\sigma$ scatter. {\bf Right Panel:} The black solid and dashed lines show the stellar mass functions of central and satellite galaxies respectively at $z=7.5$. The colored lines (see legend) show the contributions to the stellar mass functions from different halos. The error bars are $1\sigma$ poisson errors.} 
\label{SM_HM_SMF}
\end{figure*}

\begin{figure*}
\includegraphics[width=\textwidth]{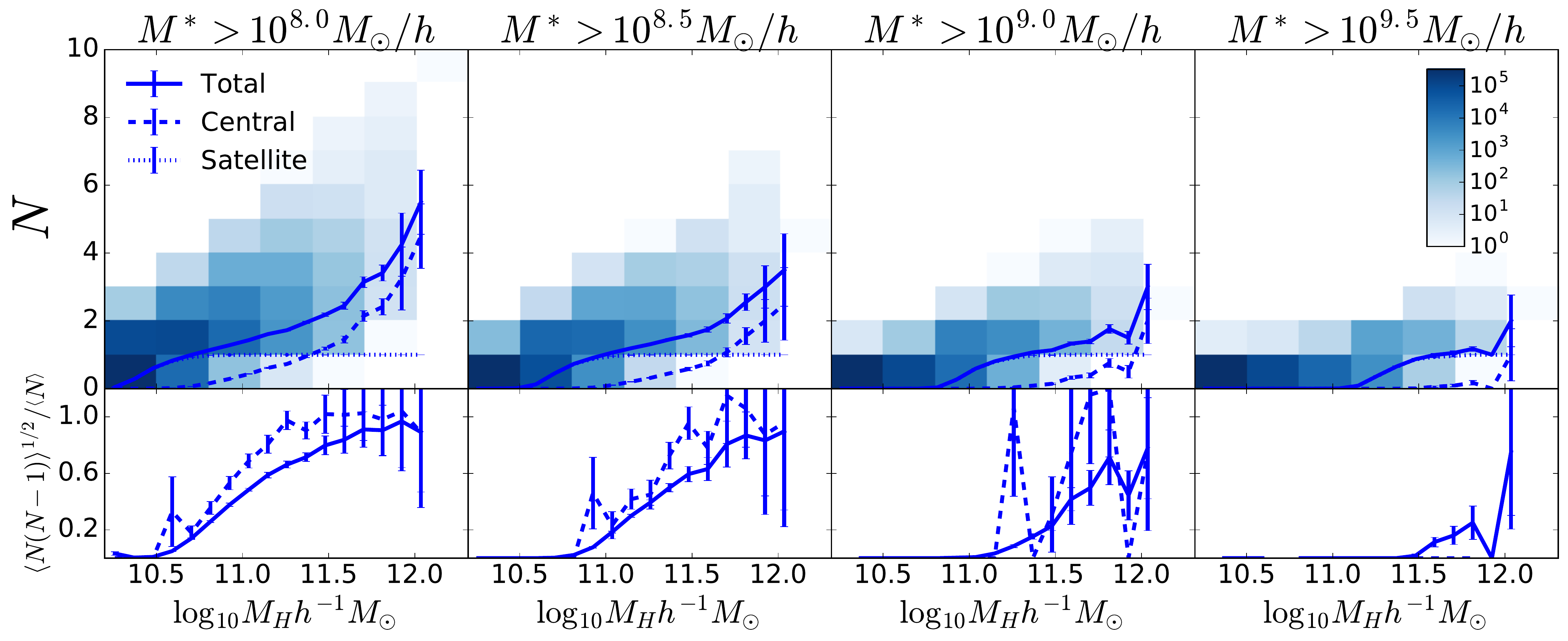}
\caption{\textbf{Upper Panels}: The blue histograms show the total Halo Occupation Distributions (HOD) for different stellar mass threshold samples in \texttt{BlueTides} at $z=7.5$. The solid lines shows the mean occupations $\left<N_{\mathrm{sat}}\right>$. The dashed and dotted lines show the mean satellite and central occupations respectively. \textbf{Lower Panels}: Solid and dashed lines show the ratio of the 2nd and 1st moments for the total halo occupations and satellite halo occupations respectively. The error bars are  computed using bootstrapping. $\left<N(N-1)\right>^{1/2}/\left<N\right> < 1$ ($>1$) implies that the distribution is narrower (broader) than a Poisson distribution.}
\label{HOD_histograms}
\end{figure*}
\label{galaxy_population}

In this work, we consider galaxy samples of stellar masses starting from $\sim10^{8} ~M_{\odot}/h$ all the way to the most massive galaxies found in \texttt{BlueTides}, $M^{*} \sim 3\times 10^{10} ~M_{\odot}/h$, build HOD models, and describe their clustering properties. 

Some example illustrations of central and satellite galaxies are shown in Figure \ref{galaxy_snaps} as red and blue points respectively on top of the dark matter distribution of the surrounding FOF group. The grey circles show the parent SO halos of the central and satellite galaxies. Figure \ref{SM_HM_SMF} (left panel) shows the stellar mass to host halo mass relation (SM-HM relation) for central galaxies and satellite galaxies at $z=7.5$. In figure \ref{SM_HM_SMF} (right panel) we also show the stellar mass function for central (dashed black lines) and satellite (solid black lines) galaxies at $z=7.5$. 
The blue, red and green solid (dashed) lines show the contributions of different halo mass bins to the the satellite (central) galaxies respectively i.e. the conditional stellar mass functions (CSMFs). For the central galaxies, the CSMFs for different halo mass bins are peaked at various stellar masses which correspond to the solid red line in figure \ref{SM_HM_SMF} (left panel). For the satellite galaxies, the CSMFs do not peak at any particular stellar mass.

Figure \ref{SM_HM_SMF} (right panel) also shows that the satellite stellar mass function is more than an order of magnitude below the central stellar mass function. Below halo masses of $3 \times 10^{10} M_{\odot}/h$, haloes do not host satellites more massive than $10^{8} M_{\odot}/h$ (which is the cut off in our analysis). Overall, we find that \texttt{BlueTides} predicts a strongly suppressed satellite population at $z>7.5$.

\subsection{Halo Occupations Distributions}

The most massive satellite galaxy formed in the simulation  at $z\sim8$ is $\sim 10^{10}M_{\odot}/h$. Accordingly, we consider the following stellar mass thresholds for our sample of satellite galaxies: $10^{8},\ 10^{8.5},\ 10^{9.0},\ 10^{9.5}~M_{\odot}/h$.

In figure \ref{HOD_histograms}, we show the full HOD for galaxies at $z=7.5$. The solid lines shows the mean occupation number. We find that even the most massive halos in \texttt{BlueTides} host a maximum of $6-7$ satellites (above the minimum stellar mass threshold). The bottom panels show the 2nd moments of the HODs; the total occupations and the satellite occupations are narrower than Poisson i.e. $\left<N(N-1)\right>^{1/2} < \left<N\right>$. The sub-poissonian nature of halo occupations have been found in previous works using semi analytical modelling \citep{2013MNRAS.435..368J}, and is somewhat expected due to low satellite occupations at these redshifts. We discuss the satellite occupation in {\tt BlueTiudes} in more detail in \S \ref{sec:satellite_HOD}.

\subsection{Mean HODs and their parametrization}

\subsubsection{Central galaxies}
\label{parametrization}

\begin{figure}
\includegraphics[width=8cm, height=6cm]{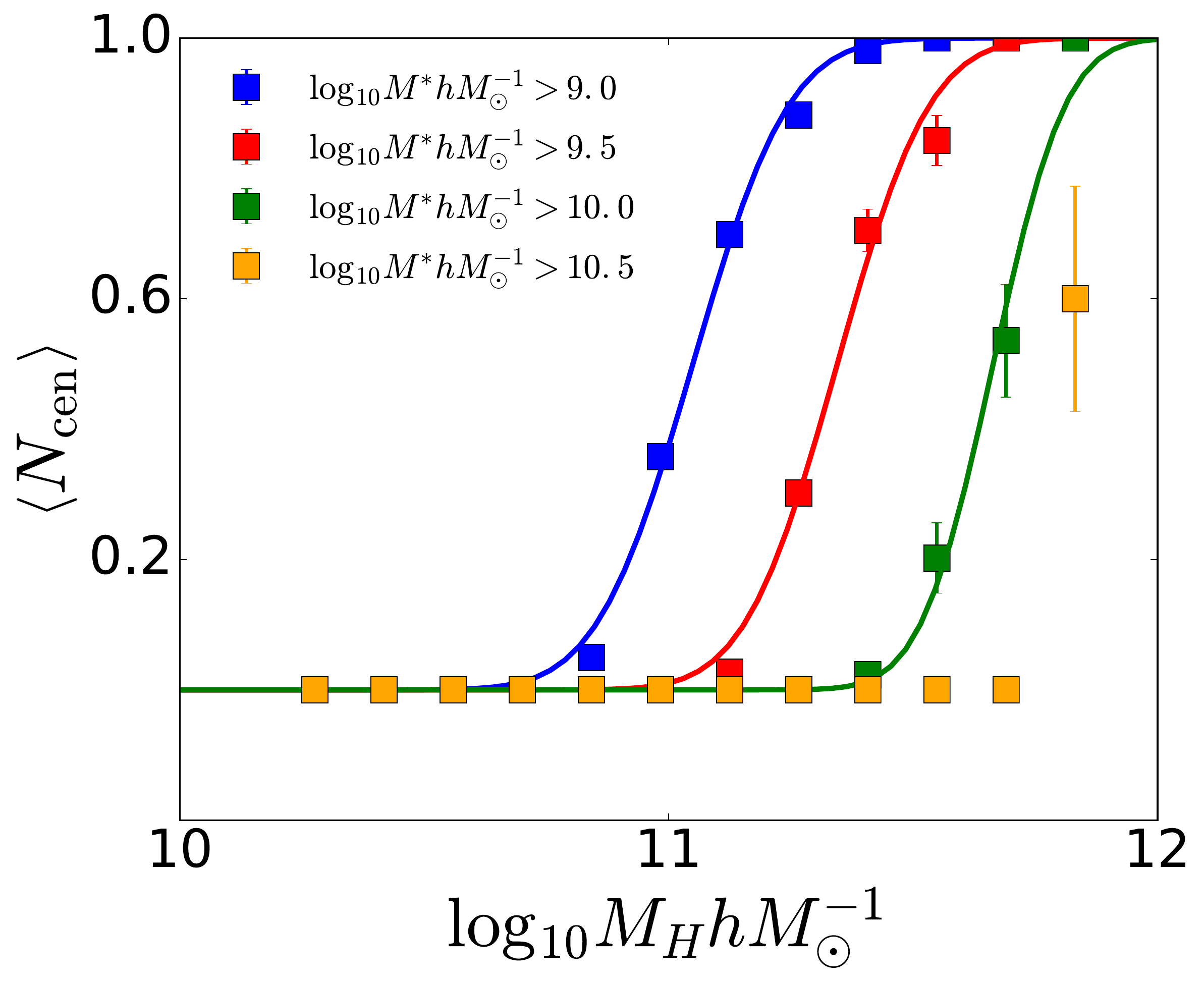}
\caption{Filled squares show the mean occupations of central galaxies for different stellar mass thresholds (see legend) at $z=7.5$. The error bars are computed using bootstrapping. The solid lines are best fits obtained for Eq. \ref{central}}
\label{mean_central_HOD}
\end{figure}

\begin{figure*}
\begin{tabular}{cc}
    \includegraphics[width=8cm, height=6cm]{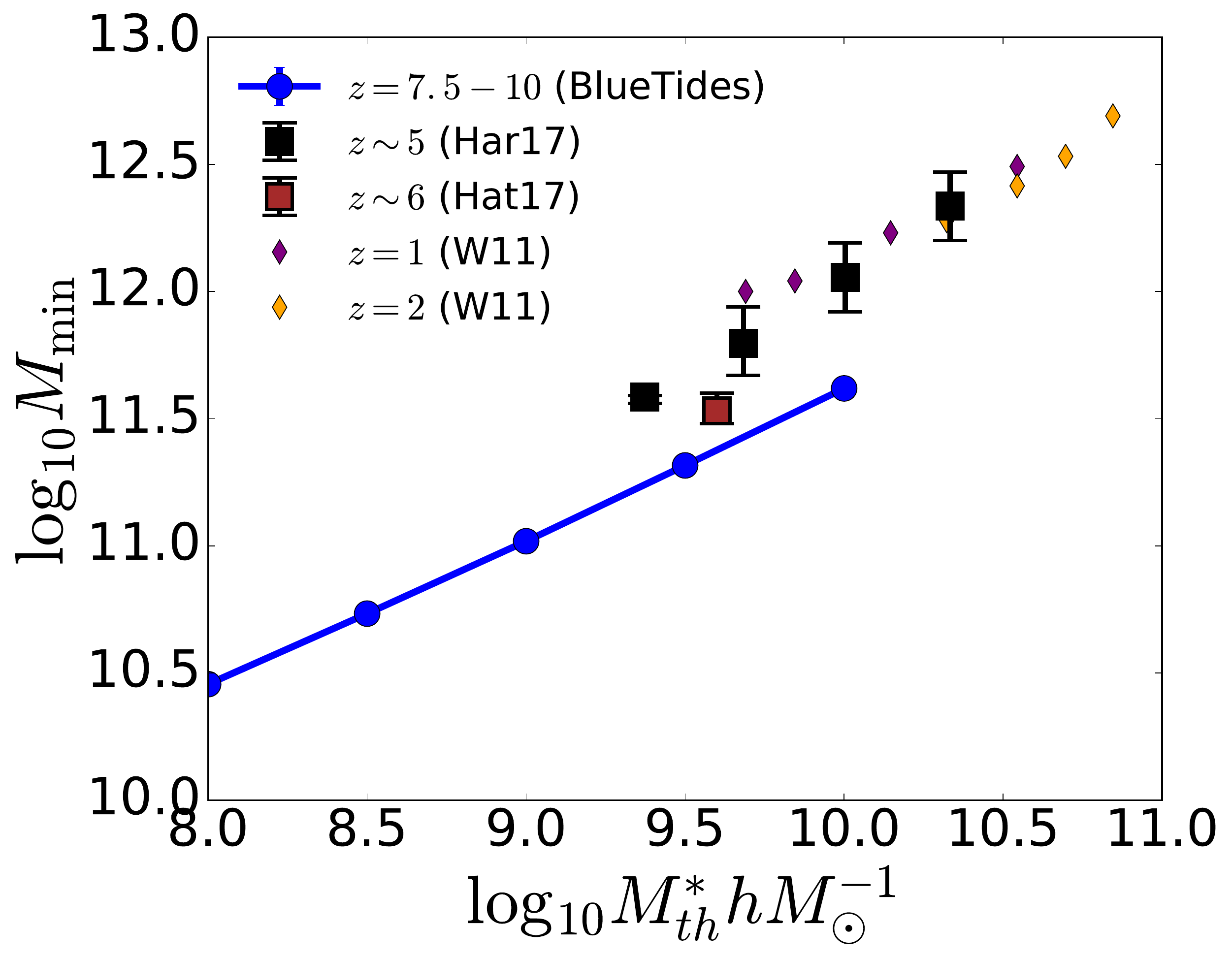}&
    \includegraphics[width=8cm, height=6cm]{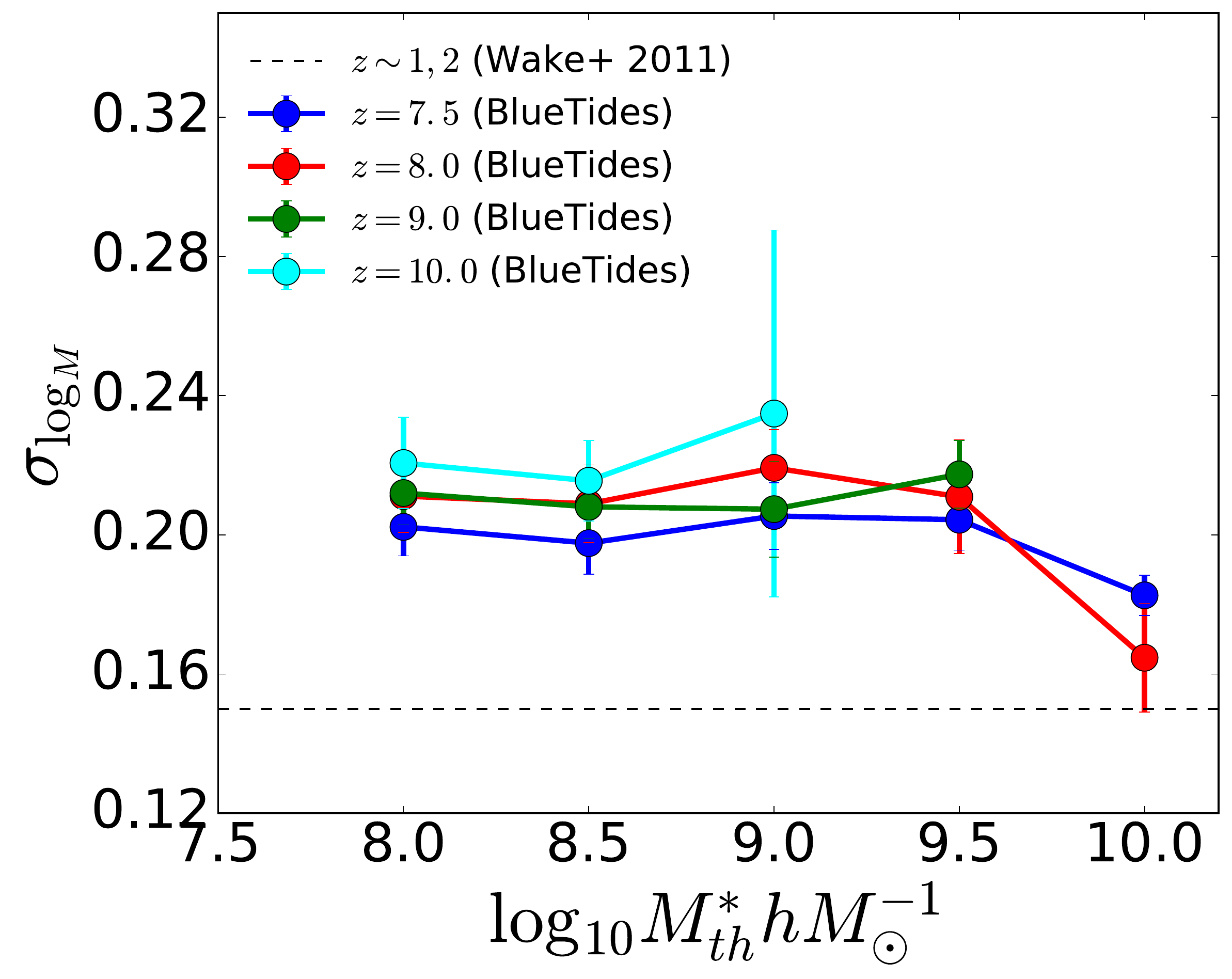}\\   
\end{tabular}
\caption{Best fit HOD parameters for mean central occupation: {\bf Left Panel:} Filled circles with solid lines show the predictions of  $M_{\mathrm{min}}$ as a function of stellar mass threshold ($M^*_{th}$). The filled squares and diamonds show the predictions from observations at low redshifts (see legend for details). {\bf Right Panel:} The filled circles with solid lines show the predictions of  $\sigma_{\log M}$ as a function of stellar mass threshold ($M^*_{th}$). The horizontal dashed line shows the value at low redshifts.  The error bars in both panels are covariance errors and are less than $1\%$ if not visible in the plot.} 
\label{mean_central_HOD_fit_params}
\end{figure*}

The filled circles in Figure \ref{mean_central_HOD} show the mean occupation of central galaxies as a function of halo mass for different stellar mass thresholds. We find that the shape of the mean central occupation does not significantly differ from HODs at low redshifts. As a result, we find that the commonly used smooth step function is a good description of the data \citep[e.g.][]{2004ApJ...609...35K, 2005ApJ...633..791Z, 2016MNRAS.463.1929M,2014MNRAS.441L..21H}
\begin{equation}
\left<N_c(M_H)\right>=\frac{1}{2}\left[1+\textrm{erf}\left(\frac{\log{M_H}-\log{M_{\textrm{min}}}}{\sigma_{\log{M}}}\right)
\right],
\label{central}
\end{equation}
where $M_{\textrm{min}}$ is the host halo mass at which $50 \%$ of halos host a central galaxy and $\sigma_{\log{M}}$ is a measure of the scatter in the stellar mass - halo mass relation. The solid lines in Figure \ref{mean_central_HOD} show the best fits for each sample. Unless stated otherwise, we fit all HOD parameters using a non-linear least squares algorithm using the {\tt Python} package \texttt{scipy.optimize.curvefit}. The values of the best fit parameters are plotted in Figure \ref{mean_central_HOD_fit_params}. The slope of the relation between $M_{\textrm{min}}$ with $M^*_{th}$ is somewhat consistent to that of observations at lower redshifts (albeit extending to lower stellar masses at high-z). When the $M_{\textrm{min}}$ values of  $z\sim 1-2$ \citep[hereafter W11]{2011ApJ...728...46W}, $z \sim 5$ \citep{2017arXiv170406535H}, and $z\sim 6$ \citep{2017arXiv170203309H} are plotted together, we find that the intercept of $M_{\textrm{min}}$ vs. $M^{*}_{\mathrm{th}}$ decreases from $z=1$ to $z=6$. While this trend is consistent with $z=7.5$, we find no further decrease in the intercept from $z=7.5$ to $z=10$. The value of $\sigma_{\log{M}}$ ($\sim 0.2-0.24$) predicted by \texttt{BlueTides} is slightly higher than the standard value of 0.15 adopted at low redshift \citep{2011ApJ...728...46W}, which implies a slightly larger scatter in the stellar mass--halo mass relation at $z \sim 8,9,10$.

\subsubsection{Satellite galaxies}
\label{sec:satellite_HOD}
\begin{figure*}
\begin{tabular}{cc}
    \includegraphics[width=8cm, height=6cm]{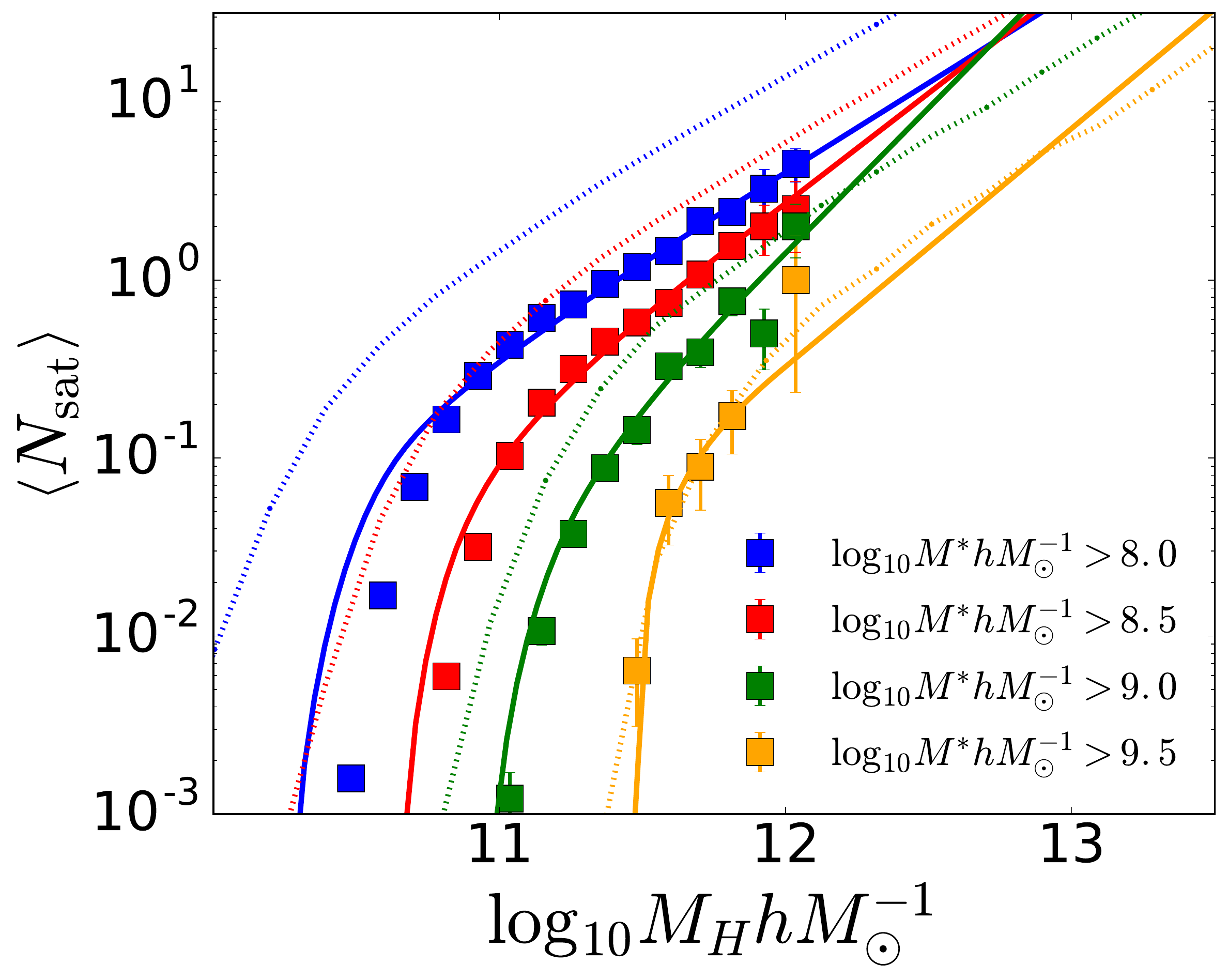}&
    \includegraphics[width=8cm, height=6cm]{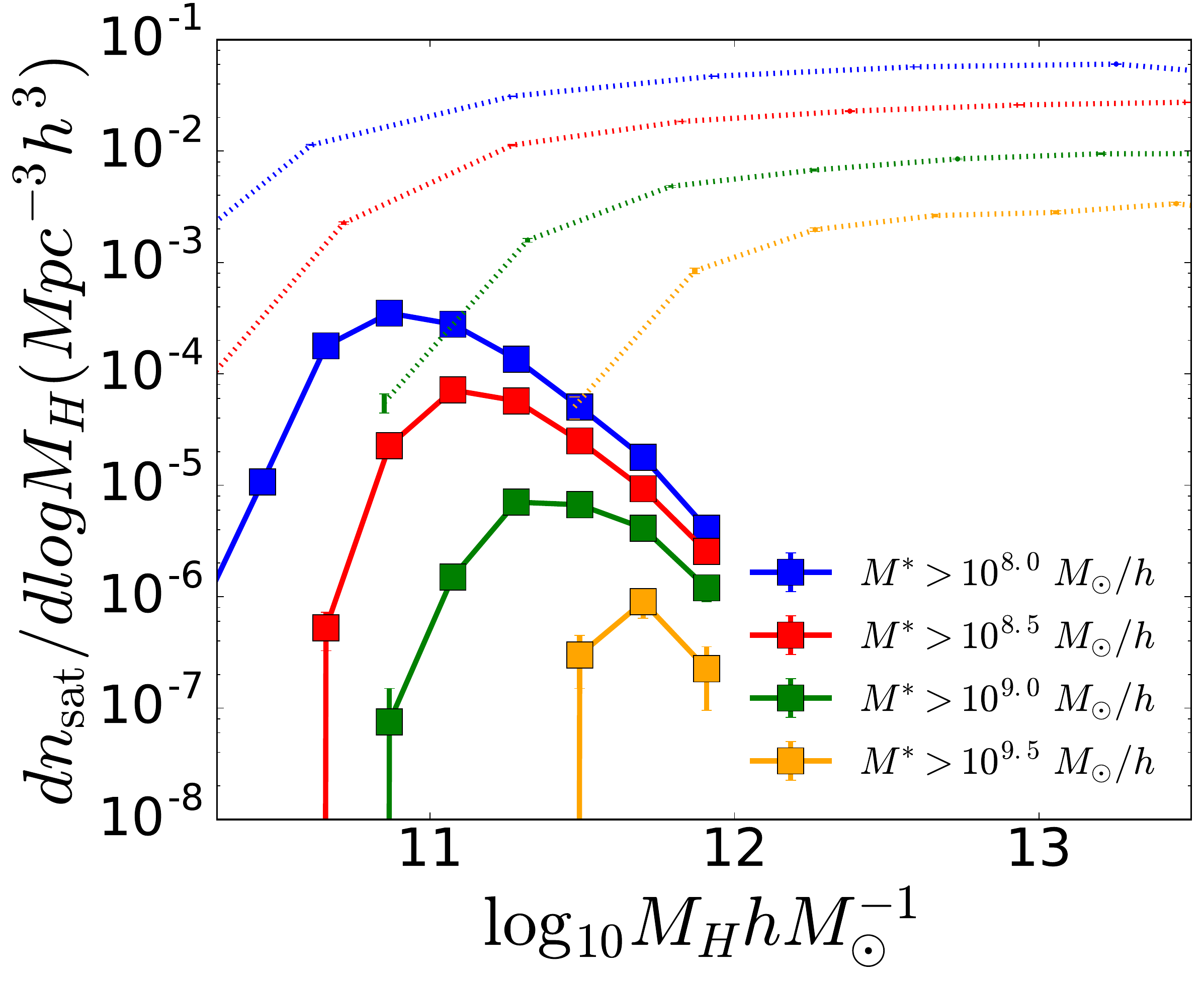}\\
\end{tabular}
\caption{{\bf Left Panel:} Filled squares show the mean occupations of satellite galaxies as a function of halo mass at $z=7.5$. The solid lines show the best fits obtained using Eq. \ref{satellite}. The error bars are computed using bootstrapping. {\bf Right Panel:} $dn_{\mathrm{sat}}/d\log{M_H}\equiv \left<N_{\mathrm{sat}}\right> dn/d\log M_H$ where $dn/d\log{M_H}$ is the halo mass function and $M_H$ is the halo mass. The dotted lines of corresponding color in both panels are predictions from \texttt{MassiveBlack II} at $z\sim0$. The errorbars are $1\sigma$ poisson errors.}
\label{mean_satellite_HODs}
\end{figure*}

\begin{figure*}
\begin{tabular}{cc}
    \includegraphics[width=8cm, height=6cm]{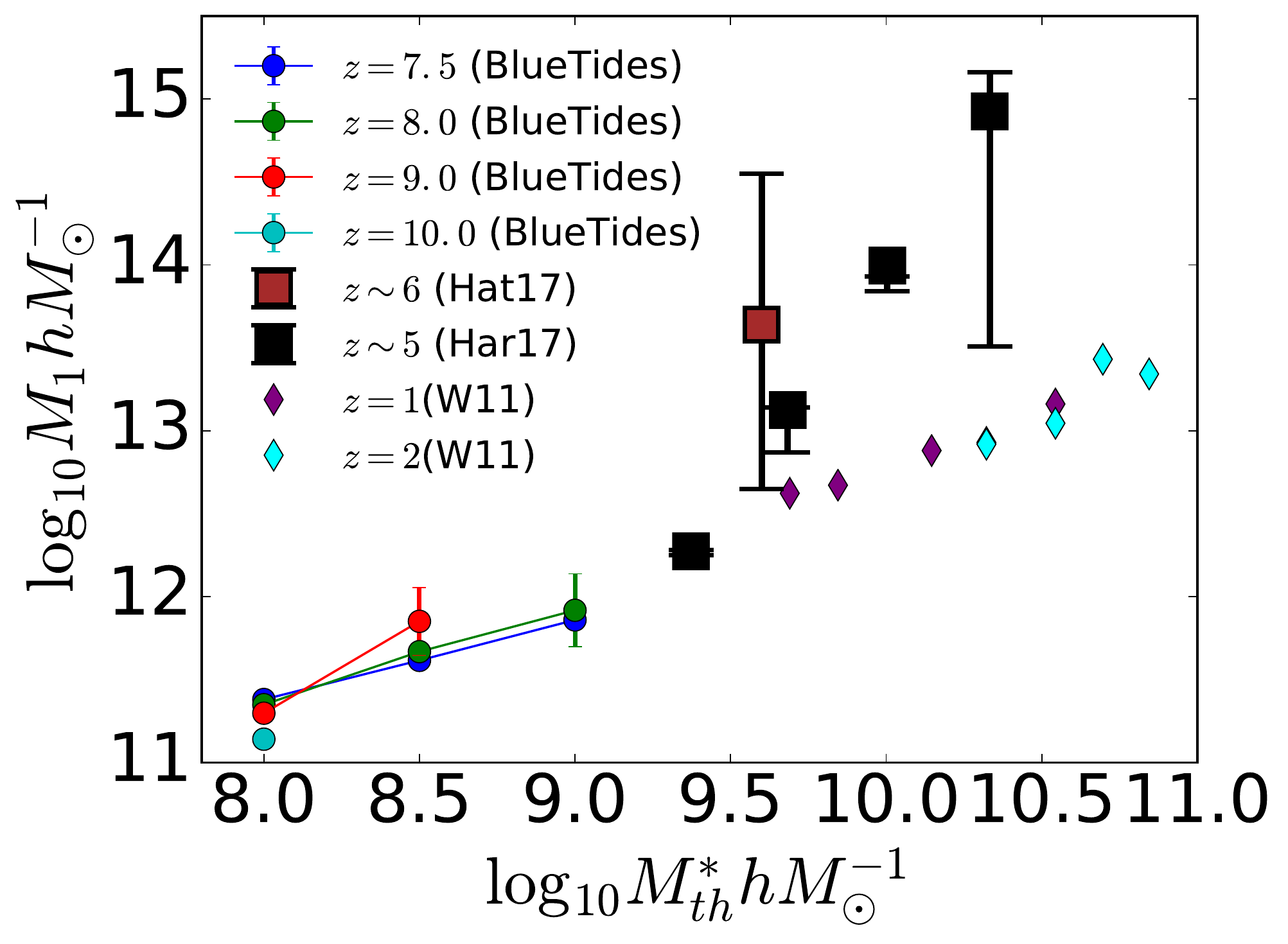}&
    \includegraphics[width=8cm, height=6cm]{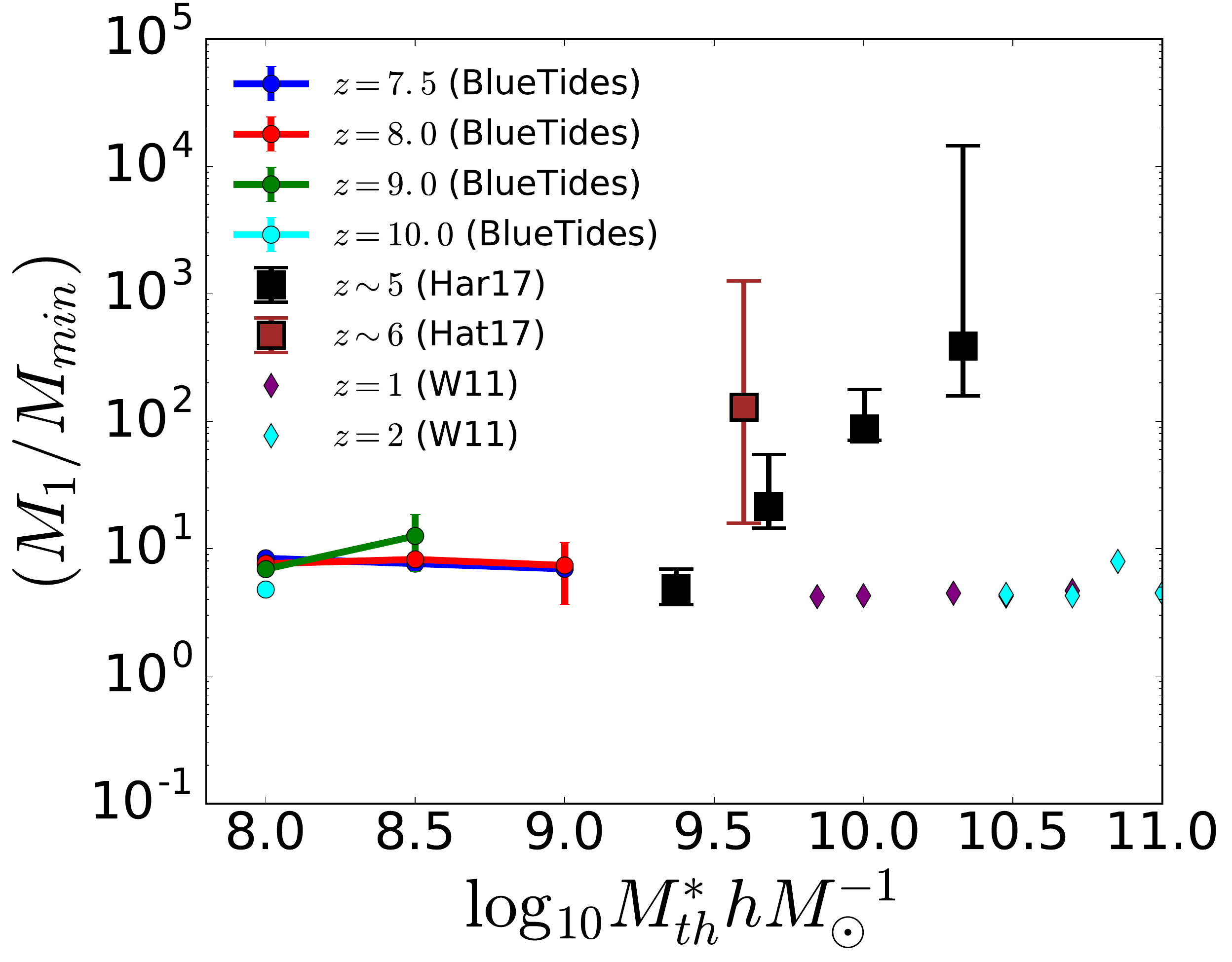}\\
\end{tabular}

\caption{Best fit HOD parameters for mean satellite occupation at different redshifts: {\bf Left Panel:} Filled circles with solid lines show the predictions of  $M_{\mathrm{1}}$ as a function of stellar mass threshold ($M^*_{th}$). {\bf Right Panel:} The filled circles with solid lines show the predictions of  $M_1/M_{\mathrm{min}}$ as a function of stellar mass threshold ($M^*_{th}$). The filled squares and diamonds in both panels show the values inferred from observations at low redshifts (see legend for details). The errorbars in both panels show the covariance errors.}

\label{mean_satellite_HODs_fit_params}
\end{figure*}

\begin{figure*}
\begin{tabular}{cc}
    \includegraphics[width=8cm, height=6cm]{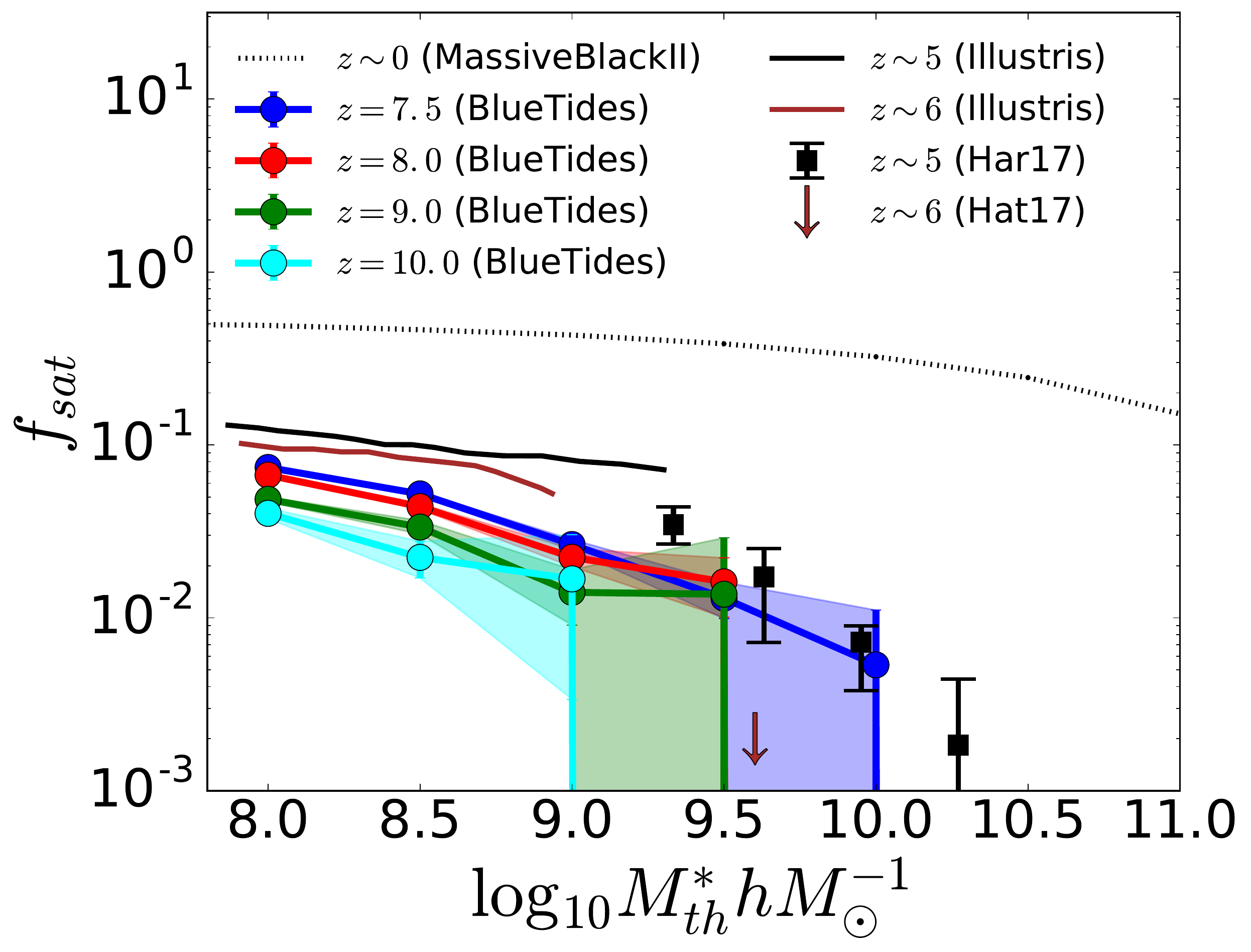}&
    \includegraphics[width=8cm, height=6cm]{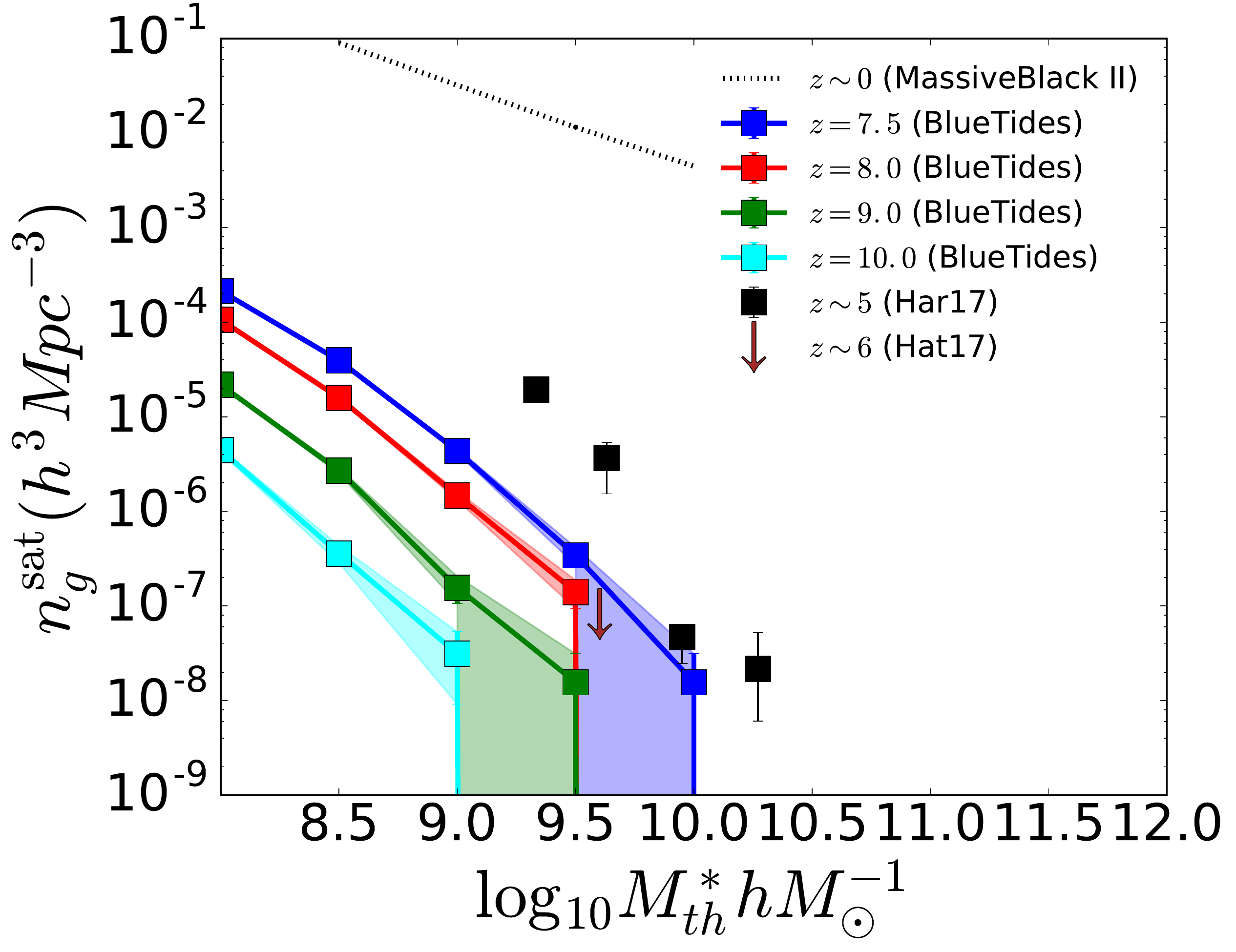}\\
\end{tabular}
\caption{Left Panel: The satellite fraction $f_{\mathrm{sat}}$ is defined as $N_{\mathrm{sat}}/N_{\mathrm{total}}$ where $N_{\mathrm{sat}}, N_{\mathrm{total}}$ are the number of satellite galaxies and total number of galaxies respectively in a given sample. The solid lines with filled circles show the predictions of \texttt{BlueTides} from $z=7.5$ to $z=10$. Black and brown solid lines are the predictions from \texttt{Illustris} at $z=5,6$. Right Panel: Filled circles with solid lines show the total number density of satellite galaxies at $z=7.5,8,9,10$. In both panels, black circles and brown arrow are observational inferences (see legend). The errorbars in both panels show $1\sigma$ poisson errors. 
\label{satellite_fraction}
}
\label{mean_HODs_thresholds}
\end{figure*}
We now turn our attention to satellite galaxies. The filled squares in Figure~\ref{mean_satellite_HODs} (left panel) shows the satellite mean occupation numbers $\left<N_{\mathrm{sat}}\right>$. $\left<N_{\mathrm{sat}}\right>$ follows a power law for large halo masses ($\gtrsim 5\times 10^{11} M_{\odot}/h$) and steeply falls off to zero eventually as the halo mass decreases. For illustration, the dotted lines of corresponding color show $\left<N_{\mathrm{sat}}\right>$ at $z\sim 0$ in the \texttt{Massive Black II} simulation. 
Noticeably the mean satellite occupation numbers are suppressed
at high redshifts compared to $z=0$      

The filled squares in Figure \ref{mean_satellite_HODs} (right panel) shows the number of satellites as a function of halo mass,  $dn_{\mathrm{sat}}/dM_{H}\equiv \left<N_{\mathrm{sat}}\right> dn/dM_{H}$. The maxima for various $M^*_{\rm th}$ shows the host halo masses that provide the dominant contribution to the satellite population at those specific stellar masses. Here the trend is somewhat different than at low redshifts (see dotted lines for \texttt{Massive Black II}): the most massive halos of e.g.; $\sim 10^{12} M_{\odot}/h$ at $z-7.5$ are exceedingly rare at $z=7.5$, so even though those are the ones that host most satellites the total number of satellites is mostly contributed by more common halos (up to a factor of $\sim 10$ lower in mass). At $z=0$, in this range of halo masses the 
number of satellites as function of halo mass remains approximately constant.

We fit the mean satellite occupations using the function 
\begin{equation}
\left<N_{\mathrm{sat}}(M_H)\right>=\left<N_{\mathrm{cen}}(M_H)\right>\left(\frac{M_H-M_0}{M_1}\right)^\alpha
\label{satellite}.
\end{equation}
where $M_0$ is the minimum cutoff mass required for a halo to host a satellite galaxy, $M_1$ is the mass scale when the halos start to host one galaxy on an average, also referred to as the `slope' of the mean satellite occupation; $\alpha$ is the power law exponent.

The solid lines show the best fit curves using Eq.~(\ref{satellite}). The filled squares in the left panel of Figure \ref{mean_satellite_HODs_fit_params} show the variation of $M_1$ as a function of $M^*_{\rm th}$ at $z\sim 7.5,8,9,10$ comparing to a range of
values infereed from a variety of observations
(as labelled on the figure).
Note that, at high$-z$ the galaxy stellar masses probed are typically $M^*_{th}<10^{9.5} M_{\odot}/h$. However, current
constraints at low redshifts are at higher masses (typically $M^*_{th}>10^{9.5} M_{\odot}/h$). Extrapolating, we find that 
the slopes of $M_1$ vs. $M_{\mathrm{th}}$ are consistent with the observational trends at $z\sim 1,2$. However, the inferred values from \cite{2017arXiv170406535H} at $z\sim 5$,  show a steeper increase in $M_1$ for $M^*_{th}>10^{9.5} M_{\odot}/h$ and would not lie on the extrapolation from 
\texttt{BlueTides}.
$M_1/M_{\mathrm{min}}$ (shown in right panel of Figure \ref{mean_satellite_HODs_fit_params}) remains roughly constant with $M^*_{th}$ for $z\sim 7.5,8,9,10$ for different stellar mass thresholds, again consistent
with inferred values at higher stellar masses (and  contrast to \cite{2017arXiv170406535H} which show a steep increase in $M_1/M_{\mathrm{min}}$ as a function of $M^*_{th}$ but these redshift and masses are not
probed by \texttt{Bluetides}).

Figure \ref{satellite_fraction} shows the satellite fraction  (defined as $f_{\mathrm{sat}}\equiv N_{\mathrm{sat}}/(N_{\mathrm{sat}}+N_{\mathrm{cen}})$)  The filled circles show the results form  \texttt{BlueTides} at $z\sim 7.5,8,9,10$. As expected, the $f_{\mathrm{sat}}$ decreases with $M^*_{\mathrm{th}}$ with the trends that appear consistent with the predictions from \texttt{ILLUSTRIS} at $z\sim5,6$. We additionally plot the total satellite galaxy densities shown in Figure \ref{satellite_fraction} (right panel). At $z=5$, the density of satellite galaxies for $M^{*}_{\mathrm{th}}=10^{10}M_{\odot}/h$ is roughly $10^{-7} h^{3}Mpc^{-3}$, which is close to the limiting value which \texttt{BlueTides} can probe given the simulation volume. The predictions do not appear to 
be in conflict with any of the current constraints 
(albeit at different redshifts).     

\section{Radial distribution of satellite galaxies}

\begin{figure*}
\centering
\includegraphics[width=18cm]{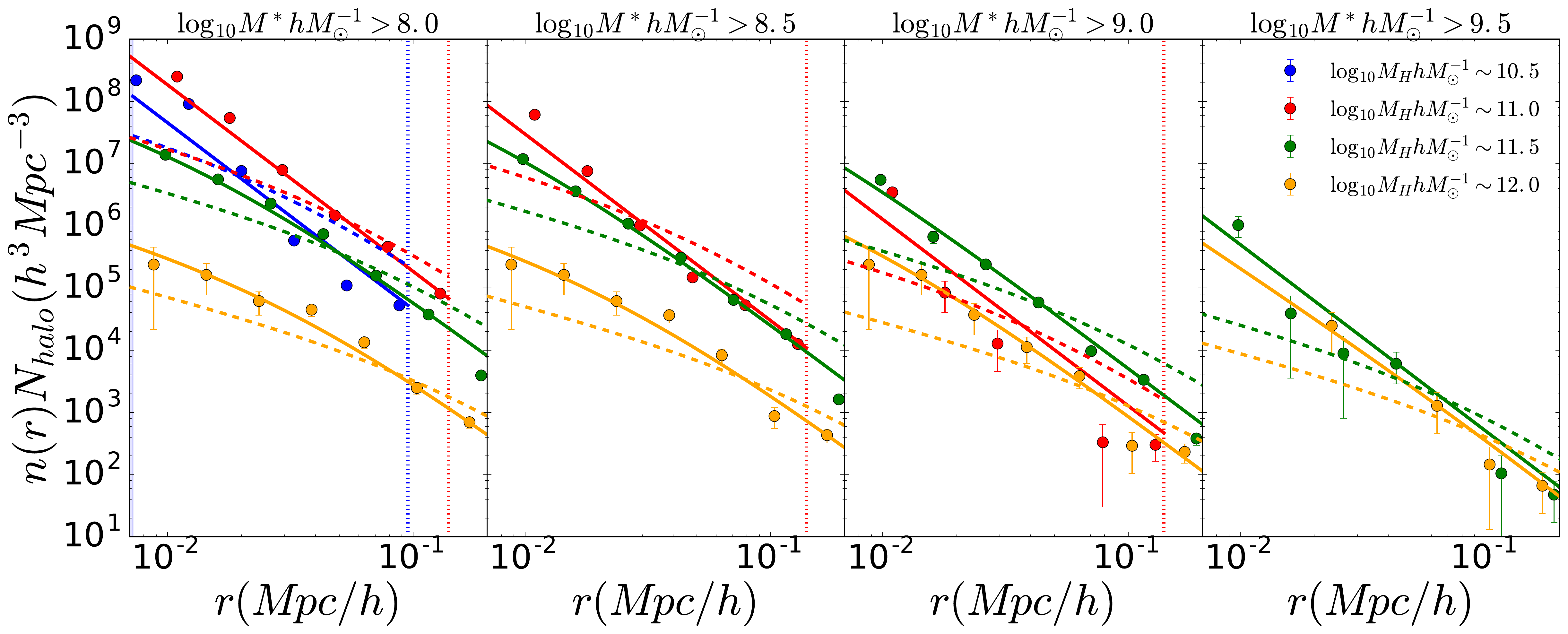}\\
\caption{$n(r)$ is the number of satellites inside a shell of radius $r$ and thickness $dr$ per unit volume per halo. The Figure shows the density profiles for $z=7.5$. $N_{\mathrm{halo}}$ is the number of halos contributing to the respective profiles in different halo mass bins (see legend). Each panel corresponds to different stellar mass threshold. The solid lines correspond to best fit NFW profile, while the dashed lines correspond to NFW profiles with concentrations from \protect\cite{2001MNRAS.321..559B}. The error bars show $1\sigma$ Poisson errors.}
\label{radial_satellite_distribution}
\end{figure*}

\begin{figure}
\includegraphics[width=8cm]{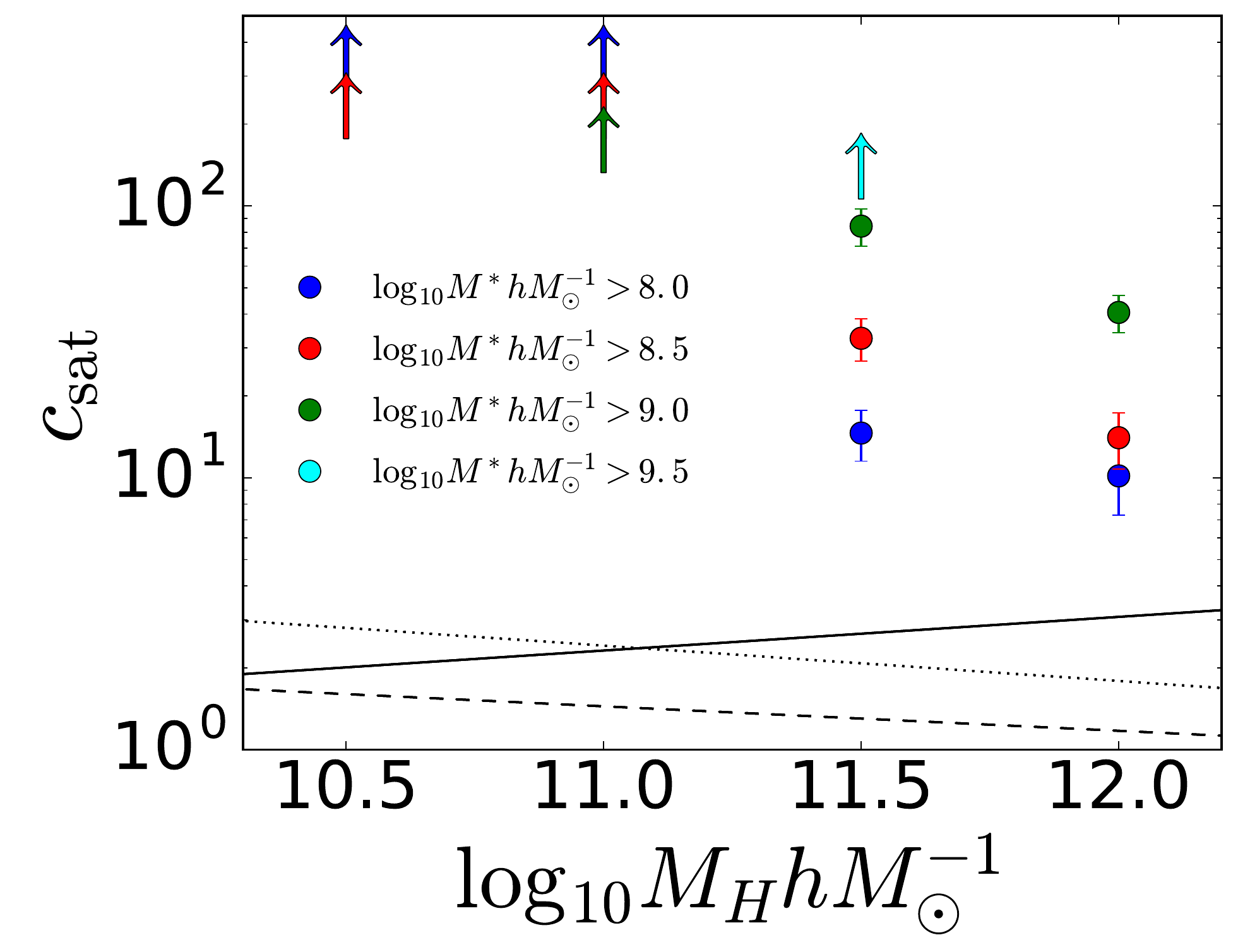}\\
\caption{Best fit concentrations obtained for the satellite density profiles using an NFW fit at $z=7.5$. The errorbars are covariance errors. The up-arrows indicate that the profiles are effectively power laws with exponent -3; $c_{\mathrm{sat}}\rightarrow \infty$ and the profiles are too highly concentrated to be constrained by an NFW fit. The error bars show the covariance errors. The black dashed and dotted lines shows the dark matter concentrations from \protect\cite{2008MNRAS.390L..64D} and \protect\cite{2001MNRAS.321..559B} respectively. The black solid line shows the dark matter concentrations of the halos (SO) in \texttt{BlueTides}.}
\label{satellite_concentrations}
\end{figure}

We now look at the radial distribution of satellites as predicted by \texttt{BlueTides} at high redshifts. The filled circles in Figure~\ref{radial_satellite_distribution} show the distribution of satellites around central galaxies at $z\sim 7.5$. Different panels correspond to different stellar mass thresholds, and the different colors in each panel correspond to different host halo mass bins; the density profiles are averaged over all halos in the corresponding halo mass bin.

We fit these profiles to a standard NFW given by
\begin{equation}
n_{\mathrm{sat}}(r)=\frac{n_{0}}{c_{\mathrm{sat}}x(1+ c_{\mathrm{sat}}x)^2}
\label{NFW_eqn}
\end{equation}
where $x=r/r_{\mathrm{200b}}$, $r_{\mathrm{200b}}$ is the radius of the host SO halo, and $200b$ refers to the fact that these halos have an overdensity of 200 times the mean density of the universe. $c_{\rm sat}$ is the satellite concentration parameter which is defined as $c_{\rm sat}=\frac{r^{\mathrm{sat}}_s}{r_{\mathrm{200b}}}$ where $r^{\mathrm{sat}}_s$ is the \textit{satellite scale radius} defined as the distance $r$ from the central galaxy at which the slope of $n_{\mathrm{sat}}(r)$ is $-2$. If $c\rightarrow \infty$, Eq.~(\ref{NFW_eqn}) effectively becomes a power law with exponent -3. The fits using Eq.~(\ref{NFW_eqn}) are shown as solid lines in Figure~\ref{radial_satellite_distribution} and the corresponding best fit concentrations are plotted in Figure~\ref{satellite_concentrations} and the values are tabulated in Table~\ref{concentration_table}. Dashed lines of corresponding color in Figure~\ref{radial_satellite_distribution} show NFWs with mass-concentrations adopted from \cite{2001MNRAS.321..559B} (which has been used to model satellites in high-z HOD modelling). We find our satellite profiles for all halo masses and stellar mass thresholds are significantly steeper than the dashed lines. 

The orange and green circles in Figure \ref{radial_satellite_distribution} show the profiles for $M_H \sim 10^{11.5} M_{\odot}/h$ and $10^{12} M_{\odot}/h$ respectively. We find that for $M^*_{th} \lesssim 10^9~M_{\odot}/h$ (first three panels from the left), the profiles in those halos are consistent with NFW (see orange and green solid lines) with concentrations $c_{\mathrm{sat}}\sim 10-40$ (see Figure~\ref{satellite_concentrations}).  Note also that the satellite concentrations in Figure~\ref{satellite_concentrations} lie significantly above the dark matter concentrations predicted by \texttt{BlueTides} (black solid line). Here we have also shown the mass concentration relations of \cite[used for modeling satellites in Har16]{2001MNRAS.321..559B} (dotted line) and \cite[used for modeling satellites in Har17]{2008MNRAS.390L..64D} (dashed line), and our satellite concentrations lie significantly above both of these lines\footnote{The solid line in Figure~\ref{satellite_concentrations} shows that dark matter concentration increases with halo mass at $z=7.5$ in \texttt{BlueTides}; this trend is consistent with the results of \cite{2015ApJ...799..108D} at $z=6$. The concentrations from \cite{2001MNRAS.321..559B} and \cite{2008MNRAS.390L..64D} show instead a decreasing concentration with halo mass, but these results were obtained at $z\sim 0$ and are extrapolated to $z = 7.5$ assuming a $(1+z)^{-1}$ scaling with redshift.}.  

For $M_H \sim 10^{10.5} M_{\odot}/h$ and $10^{11} M_{\odot}/h$ (blue and red circles in Figure \ref{radial_satellite_distribution}), the satellite profiles for all stellar mass thresholds cease to be NFW and become a power law with slope -3 (equivalently a limiting case of NFW with $c_{\mathrm{sat}}\rightarrow \infty$). These are shown as up-arrows in Figure~\ref{satellite_concentrations}. Satellites with $M_H \gtrsim 10^{9.5} M_{\odot}/h$ also trace a power law profile with slope -3; however these satellites are exceedingly rare and virtually impossible to detect in current as well as future surveys (as we shall see in section \ref{probabilities}).

Below, we summarize our key findings about the satellite density profiles:
\begin{itemize} 
\item For halos with $M_H \gtrsim 3\times 10^{11}~M_{\odot}/h$, satellites with $10^8 \lesssim M^*\lesssim 10^{9} M_{\odot}/h$ are consistent with NFW but the concentrations are higher ($c_{\mathrm{sat}}\sim 10-40$) than dark matter concentrations used to describe satellite profiles in recent works on high-z HOD modeling (Hat17, Har17).


\item  For halos with $M_H \lesssim 3\times 10^{11}~M_{\odot}/h$, satellites profiles are a power law with slope -3 for all $M^* \gtrsim 10^8~M_{\odot}/h$.
\end{itemize}
The lower mass halos ($M_H \lesssim 3\times 10^{11}~M_{\odot}/h$) tend to dominate the satellite population. We just established that satellite profiles in these halos are a power law. This also implies that satellite galaxies with $M^*\gtrsim 10^8~M_{\odot}/h$ are poor tracers of dark matter at these redshifts. This is not surprising given that individual dark matter halos at these redshifts mostly host between 0 and 1 satellites.    

In the next section, we investigate the implications on the small scale clustering at these redshifts.

\section{Implications on galaxy clustering}

\begin{figure*}
\includegraphics[width=18cm]{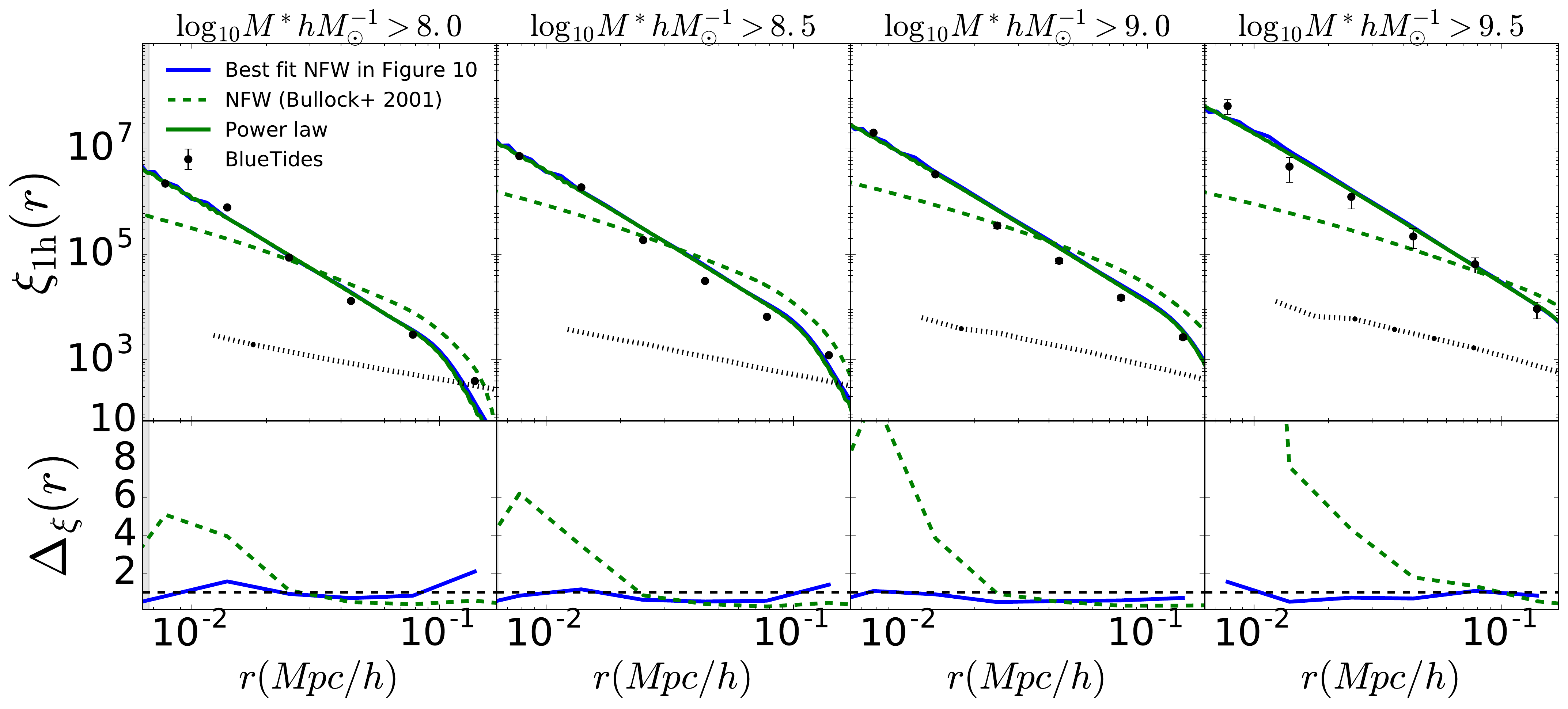}
\caption{{\bf Upper Panels:} One-halo clustering predicted by \texttt{BlueTides} for various stellar mass thresholds. The solid blue line is the HOD model prediction using the best fit NFWs for the satellite densities shown in Figure \ref{radial_satellite_distribution}. The dashed green line is the HOD model prediction using NFW with concentrations from \protect\cite{2001MNRAS.321..559B}. The solid green line shows the HOD model prediction using a power law profile with slope -3. {\bf Bottom Panels:} The solid and dashed lines of corresponding color (see top panels) shows the ratio between the simulations and the HOD model predictions. The error bars show the $1\sigma$ Poisson errors.}
\label{one_halo_clustering}
\end{figure*}

\begin{figure}
\includegraphics[width=8cm]{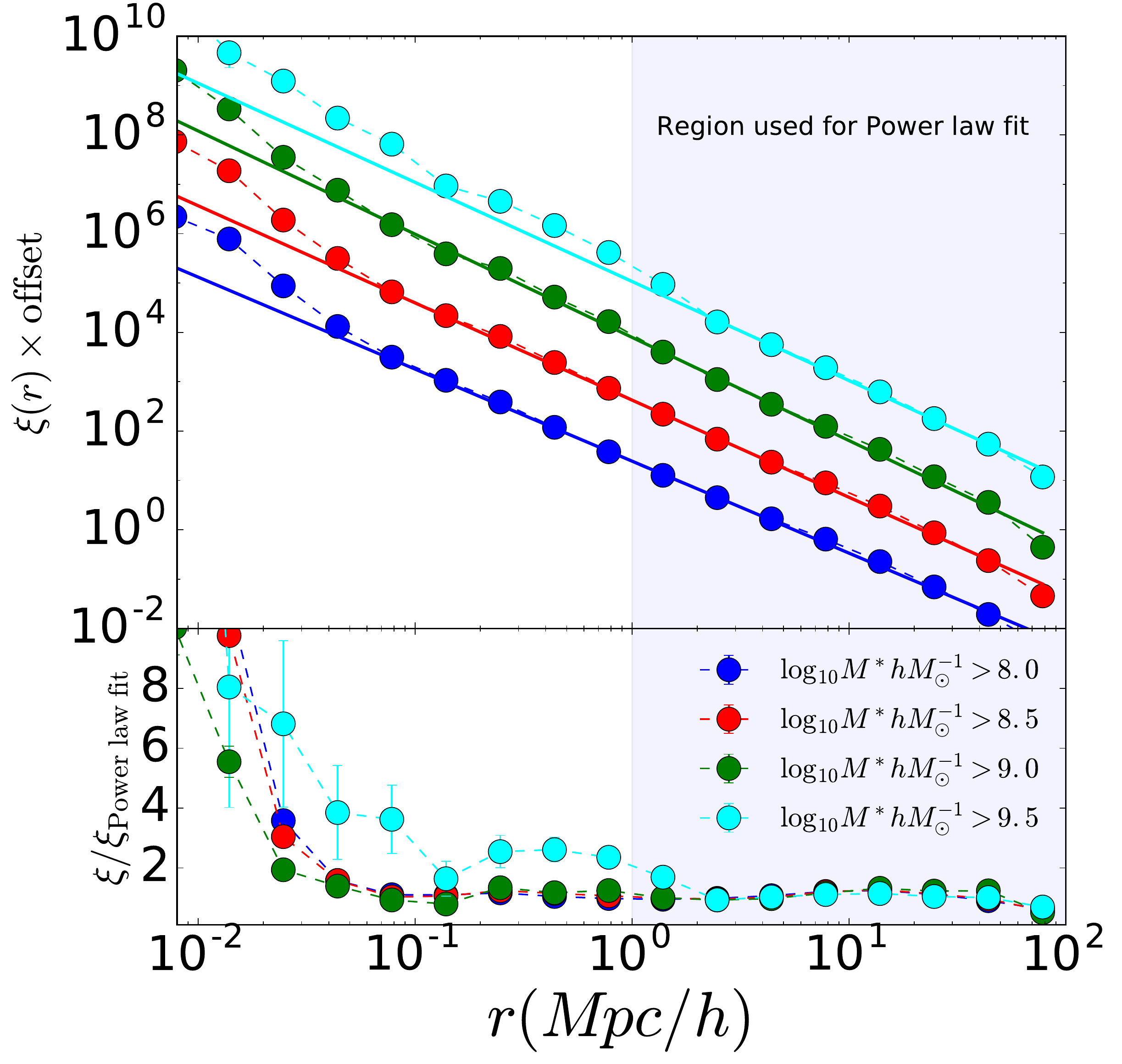}
\caption{Filled circles show the total clustering profile of all galaxies for different stellar mass thresholds predicted by \texttt{BlueTides} at $z=7.5$. The solid lines show power law fits performed using the data points at large scales ($r>1~\mathrm{Mpc}/h$). The bottom panels show the ratio, i.e. the deviation of the simulation predictions from power law. The error bars show the $1\sigma$ Poisson errors.} 
\label{power_law_fit}
\end{figure}

\subsection{Small scale clustering}
\label{galaxy_clustering}

We now look at the one-halo scale clustering of high redshift galaxies and the implications of our findings on the modeling of one-halo clustering. The filled circles in Figure \ref{one_halo_clustering} show the one-halo clustering predictions of \texttt{BlueTides} for the stellar mass thresholds considered in the previous section. For reference, we also show the clustering signal for these stellar mass thresholds at $z=0$, as predicted by the \texttt{MassiveBlack II} simulation; the one halo clustering at high redshifts is much more enhanced (by over two orders of magnitude at $r \sim 10^{-2}~{\rm Mpc}/h$) than those at low redshifts. This reflects how highly biased these high-z galaxies are compared to low-z galaxies of similar stellar masses.  

\subsubsection{HOD modelling of the one-halo term}
The green dashed line in Figure \ref{one_halo_clustering} shows the HOD model predictions with a mass-concentration relation from \cite{2001MNRAS.321..559B}; as we can see our simulations predict a significantly steeper profile. The solid blue line shows the HOD model predictions obtained from the fits to satellite profiles shown in Figure \ref{radial_satellite_distribution}, and shows significant improvement compared to the green dashed lines. In the previous section, we found that the halos which predominantly host satellite galaxies ($M_H\lesssim 3 \times 10^{11} M_{\odot}/h$) have satellite density profiles which are power-law with slope -3. We therefore make an HOD model prediction with such a profile shown as the solid green line in Figure \ref{radial_satellite_distribution}. The solid green and solid blue lines agree. This tells us that satellite galaxies at these high redshifts may be simply modeled by a power law with slope -3.

\subsection{Total clustering: How close is it to Power law?}

Given that we find so few satellites at these high redshifts, it would also be interesting to see if there are any potential implications on the overall clustering profile (one-halo+two-halo term). Figure \ref{power_law_fit} shows the total clustering profile for different stellar mass thresholds at $z\sim7.5$. We perform a power law fit to the large scale clustering from scales of 1 Mpc/$h$ to 100 Mpc/$h$ and look for deviations from power-law at smaller scales ($r<1~\mathrm{Mpc}/h$). We note that the power-law behavior prevails all the way to ($r<0.1~\mathrm{Mpc}/h$), which is a direct consequence of the non-linear halo bias in the two-halo regime \citep{2017MNRAS.469.4428J,2016MNRAS.463..270J};\texttt{BlueTides} clustering confirms that non-linear bias is indeed required to model clustering at quasilinear scales ($0.5-10 \mathrm{Mpc}/h$) as seen in Figure 5 of our earlier paper \citep{2018MNRAS.474.5393B}. In the one-halo regime ($r<0.1~\mathrm{Mpc}/h$), simulations predict enhancement (by factors of upto $\sim10$) compared to power law. At first glance, this may seem counterintuitive because in case of low satellite fractions, one may expect close to power-law behavior or a decrement (lower than power-law). However, detailed study on this aspect performed by \cite{2011ApJ...738...22W} revealed that these enhancements may very well occur despite very low satellite fractions if the galaxies are themselves very rare (overall number density is very low), which is indeed the case at high redshifts. Thus, given the overall number densities, one would require even lower satellite fractions for the power-law behavior to persist below 0.1 Mpc/$h$

\section{Implications for finding high redshift satellite galaxies in future surveys}
\label{probabilities}
\begin{figure*}
\begin{tabular}{cc}
    \includegraphics[width=8cm, height=6cm]{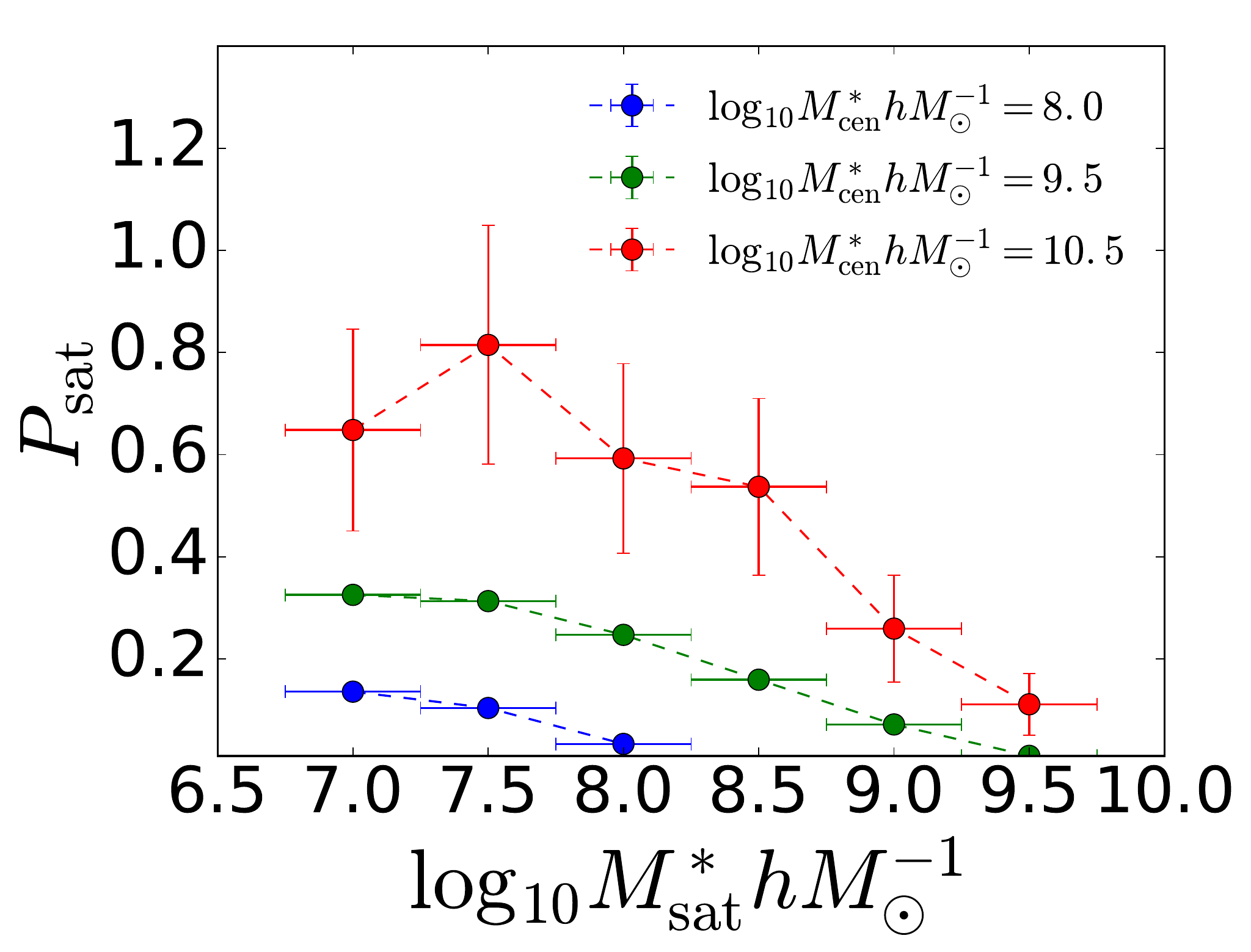}&
    \includegraphics[width=8cm, height=6.5cm]{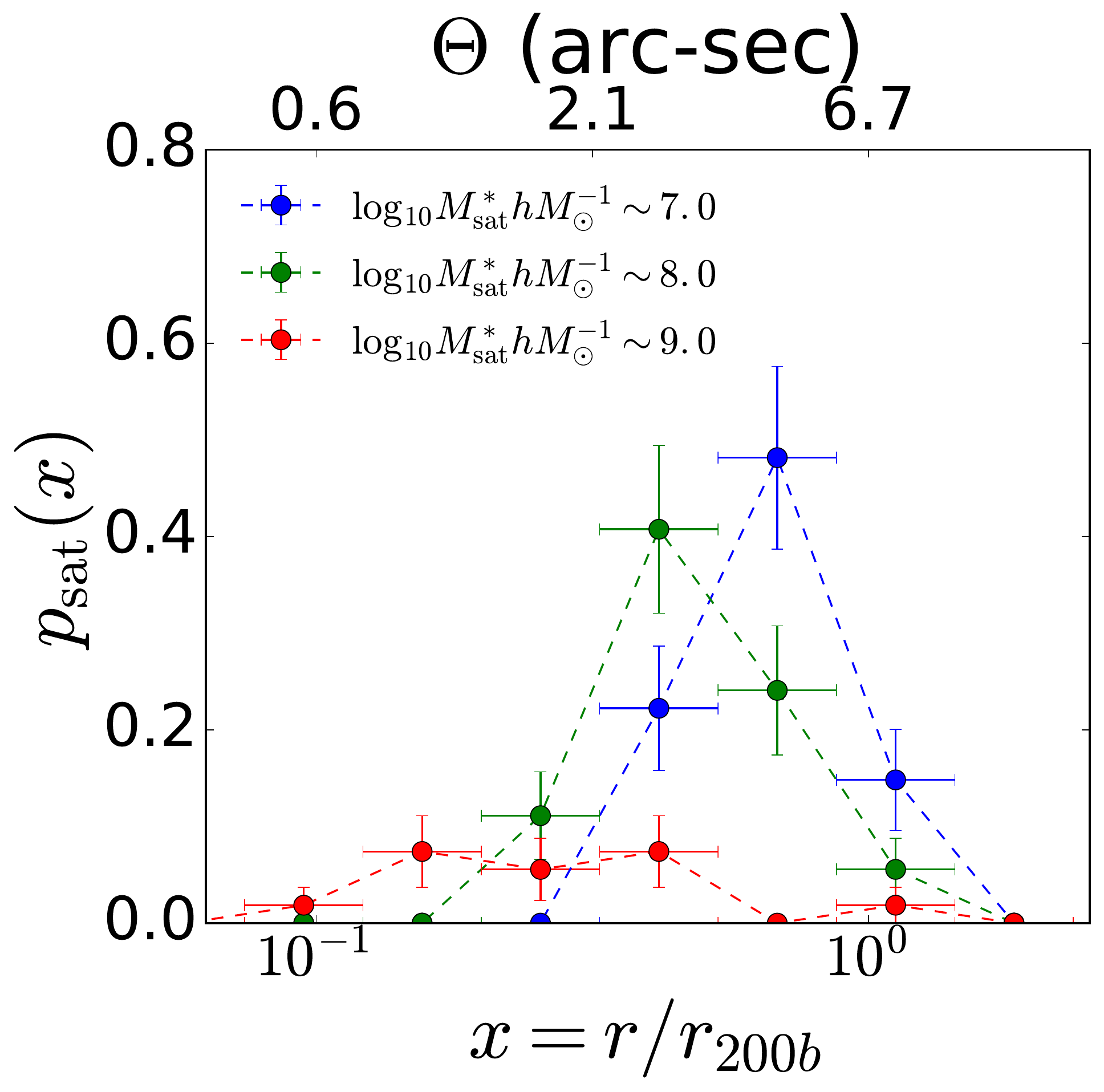}\\ 
\end{tabular}
\caption{\textbf{Left Panel}: $P_{\rm sat}$ is the probability of a central galaxy of a given mass $M^*_{\mathrm{cen}}$ to host at least one satellite galaxy of mass $M^*_{\mathrm{sat}}$ in its vicinity at $z=7.5$ as predicted by \texttt{BlueTides}. Different colors correspond to different central galaxy mass $M^*_{\mathrm{cen}}$ (see legend). The errorbars along y-axis show $1\sigma$ poisson errors, and the error bars along x-axis show the chosen bin-width of $M^*_{\mathrm{sat}}$. \textbf{Right Panel}: $p_{\rm sat} (x)$ is the conditional probability of finding at least one satellite galaxy of mass $M^*_{\mathrm{sat}}$ located at a distance $x\equiv r/r_{200b}$ from the central galaxy of mass  $M^*_{\mathrm{cen}} \sim 3\times 10^{10} M_{\odot}/h$, at $z=7.5$ as predicted by \texttt{BlueTides}. The bin widths chosen for both central and satellite stellar masses are $\delta \log_{10} M^*_{\mathrm{cen/sat}}= 0.5$ dex. Different colors in each panel correspond to different satellite galaxy masses (see legend). The errorbars along y-axis show $1\sigma$ poisson errors, and the error bars along x-axis show the radial bin widths. The scales shown in this plot is well within the fields of view of HUDF ($\sim 5 Mpc/h$) and JWST-NIRCAM ($\sim 10 Mpc/h$) at $z\sim 7.5$}
\label{probability}
\end{figure*}

Given the paucity of satellites in \texttt{BlueTides} at $z \gtrsim 7.5$ (consistent with what is inferred from observations at $z\sim 5,6$), finding satellites and probing the one-halo term may prove to be difficult/ expensive without an informed strategy. In this section, we shall use the \texttt{BlueTides} data and compute probabilities of finding satellites under different conditions, and suggest approaches to maximize the likelihood of finding satellites based on \texttt{BlueTides}. A somewhat similar study \citep{2018ApJ...856...81R} was performed very recently at $z \gtrsim 7.5$ using the conditional luminosity function (CLF) formalism. Their focus was however completely on central galaxies, whereas we focus on finding the probability of detecting satellites around centrals. $P_{\mathrm{sat}}$ plotted in Figure \ref{probability} (left panel) for $z=7.5$, is the overall probability of a central galaxy of mass $M^*_{\mathrm{cen}}$ to host at least one satellite galaxy of mass $M^*_{\mathrm{sat}}$. Different colors correspond to different central galaxy masses ranging from the detection limits in the deepest fields of HST ($\sim 10^{8} M_{\odot}/h$: blue color) to the most massive galaxies found in \texttt{BlueTides} ($\sim 3\times 10^{10} M_{\odot}/h$: red color). We look at satellite galaxy masses going all the way to the  detection limits of JWST i.e. $M_{UV} \lesssim -16$ \citep{2011MNRAS.414..847S} converted to $M^*\gtrsim 10^{7} M_{\odot}/h$ based on \texttt{BlueTides}. The central galaxies of mass $M^*_{\mathrm{cen}}\sim 10^{8} M_{\odot}/h$ have a low probability of having satellite pairs ($\sim 15 \%$ for $M^*_{\mathrm{sat}} \sim 10^{7} M_{\odot}/h$). The probability increases as the central galaxy mass increases, and for the most massive centrals $M^*_{\mathrm{cen}}\sim 3\times 10^{10} M_{\odot}/h$, there is $\sim 60-80 \%$ probability of finding satellites at the edge of the JWST detection limit ($M^*_{\mathrm{sat}} \sim 10^{7} M_{\odot}/h$). The most massive satellites ($M^*_{\mathrm{sat}} \sim 3 \times 10^{9} M_{\odot}/h$) have $\sim 10 \%$ detection probability around the most massive centrals ($M^*_{\mathrm{sat}} \sim 3\times 10^{10} M_{\odot}/h$). 

It is also interesting to look at the detection probabilities as a function of distance, particularly around the most massive centrals ($M^*_{\mathrm{cen}}\sim 3\times10^{10}\ M_{\odot}/h$) where we are most likely to find satellites. $p_{\rm{sat}} (x)$ plotted in Figure \ref{probability} (right panel) at $z=7.5$, is the conditional probability of finding at least one satellite of mass $M^*_{\mathrm{sat}}$ at a distance $x\equiv r/r_{200b}$ from a central galaxy of mass $M^*_{\mathrm{cen}}\sim 3\times10^{10}\ M_{\odot}/h$. Note that the scales shown are well within fields of view of HST-HUDF ($\sim 5$ Mpc at $z=7.5$) and JWST-NIRCAM ($\sim 10$ Mpc at $z=7.5$). For the lowest mass satellites $M^*_{\mathrm{sat}}\sim 10^{7} M_{\odot}/h$, we find that the probability is highest at a distance $\sim 100$ kpc, which is significantly far away from the satellite galaxies. As we go closer to the central galaxy, more massive satellites start to dominate; for example, at $r\sim 10-20$ kpc, the only galaxies that exist are $M^*_{\mathrm{sat}}\sim 10^{9} M_{\odot}/h$ (red circles). This is consistent with our findings in the radial density profiles of satellites in Figure \ref{radial_satellite_distribution} and \ref{satellite_concentrations}, which suggested that satellites become more centrally concentrated with increase in stellar mass.

We can therefore infer that one possible strategy to optimize the search for satellite galaxies at $z>7.5$ using JWST is to first use the largest possible field of view (e.g.:- JWST-NIRCAM with $2 \times 2.2'$ field of view) to find as many massive central galaxies ($M^*_{\mathrm{cen}}\gtrsim 10^{10}~M_{\odot}/h$) as possible. One can then find satellites around the vicinity of these massive centrals upto angular scales of $\sim 5''$ (corresponding to $\sim 0.2~{\rm Mpc}/h$ at $z\sim7.5$). Therefore, despite the obvious lack of satellites at high-z, JWST will still be able to find satellites (particularly $M^*_{\mathrm{sat}}\sim 10^{7}~M_{\odot}/h$ with $80 \%$ likelihood) to probe the one-halo regime.   

\section{Summary and conclusions}
\label{conclusion}
We analysed the high redshift galaxy population ($M^*\gtrsim 10^8 M_{\odot}/h$) in the \texttt{BlueTides} simulation, and constructed HOD models for galaxies of various stellar masses between $z=7.5-10$ that are able to successfully describe how galaxies cluster on small scales.  One of our primary findings is that standard HOD parameterizations used in low redshift studies are sufficient for modelling these galaxies.

With respect to the HODs, we find that, similar to lower redshift parameterizations, the mean number of central galaxies as a function of halo mass is well described by a smoothened step function with zero galaxies at low halo masses and one at high masses.  For satellites, the mean number follow an increasing power law function at high halo masses with a steep cutt-off at low halo masses.  Furthermore, we find no significant evolution in the HOD between $z \sim 7.5-10$ in {\tt BlueTudes}.

A significant, but not unexpected finding, is that satellite galaxies are exceedingly rare at these high redshifts.  The most massive halo in \texttt{BlueTides} at $z=7.5$ hosts 10 satellites with a stellar mass, $M_* \geq 10^{8.0}~M_{\odot}/h$.  The number of satellites drops steeply with increasing redshift and stellar mass threshold.  For halo masses below $3 \times 10^{11} M_{\odot}/h$, the mean satellite occupation $\left<N_{\mathrm{sat}}\right> <1$. $\left<N_{\mathrm{sat}}\right> \sim 0.1$ for halos that contribute the most to the satellite population. 
The satellite fractions at $z \gtrsim 7.5$ found in \texttt{BlueTides} range between $8\%$ for the $M^*_{th}\sim 10^{8} M_{\odot}/h$ to $\lesssim 1\%$ for $M^*_{th}\sim 3 \times 10^{9} M_{\odot}/h$; there are no higher mass satellites. These trends are consistent with very low satellite fractions inferred from observations at $z\sim5,6$. 

For the radial number density profiles, the satellites with $10^8 \lesssim M^*\lesssim 10^{9} M_{\odot}/h$ in halos with $M_H \gtrsim 3 \times10^{11} M_{\odot}/h$ are consistent with NFW, but the concentrations are higher ($c_{\mathrm{sat}}\sim 10-40$) than standard dark matter concentrations used to describe satellite profiles in recent works on high-z HOD modeling. Satellites inside halos with $M_H \lesssim 3 \times 10^{11} M_{\odot}/h$ follow a power law profile with slope -3. These halos dominate the satellite population and the small scale clustering, therefore only after incorporating their satellite profiles in the HOD model are we able to reproduce the simulated small scale clustering seen in the \texttt{BlueTides} simulation.
 
Lastly, the paucity of $M^*\gtrsim 10^{8} M_{\odot}/h$ satellites does not defeat the possibility of finding satellite galaxies and probing the one-halo term at these redshifts. Given the detection limits of JWST ($\sim 10^{7} M{\odot}/h$), we find that the probability of finding satellites can be boosted to $\sim 80 \%$ for a $M^*\sim 1-3 \times 10^{7} M{\odot}/h$ around the vicinity (within $\sim 7``$ of angular separation in the sky) of centrals with $M^{*}\gtrsim 3\times 10^{10} M_{\odot}/h$.
 
\section{Acknowledgements}
We acknowledge funding from NSF
ACI-1614853, NSF AST-1517593, NSF AST-1616168, NSF AST-1716131,
NASA ATP NNX17AK56G and NASA ATP 17-0123
and the BlueWaters PAID program. The \texttt{BLUETIDES} simulation was run on the BlueWaters facility at the National Center for Supercomputing Applications. AKB is thankful to Peter Behroozi, Francois Lanusse and Yao Yuan Mao for valuable inputs and discussions.
DC is supported by a McWilliams Postdoctoral Fellowship.

\appendix
\section{BlueTides: Halo mass function}
\label{mass_function}
\begin{figure}
\includegraphics[width=8cm]{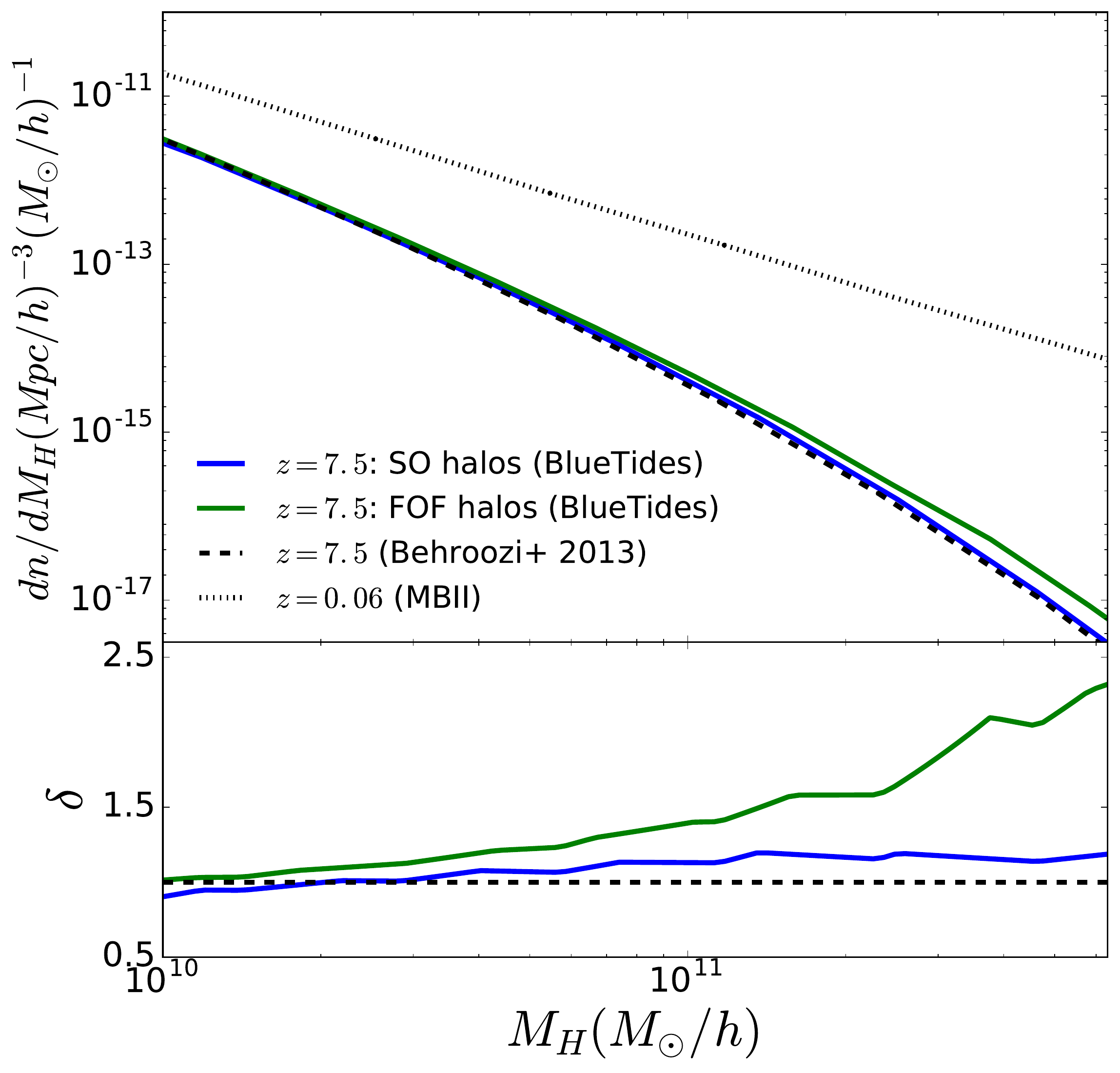}
\caption{\textbf{Top Panel}: Solid lines show the halo mass functions at $z=7.5$ using both FOF (green) and SO halos (blue) as predicted by \texttt{BlueTides}. Black dashed line is the analytical fit from \protect\cite{2013ApJ...770...57B}. The dotted grey line is the halo mass function at $z=0$ as predicted by \texttt{MassiveBlack II}. \textbf{Bottom Panel}: $\delta$ is the ratio of the mass functions at $z=7.5$ w.r.t. the analytical fit from \protect\cite{2013ApJ...770...57B}.} 
\label{mass_function_fig}
\end{figure}
We have generated both FOF (with $\texttt{LINKING_LENGTH}=0.2$ times average particle spacing in the simulation) and SO (200 times mean matter density) halo catalogs for the \texttt{BlueTides} snapshots using FOF finder \citep{1985ApJ...292..371D} and \texttt{ROCKSTAR} \citep{2013ApJ...770...57B} halo/subhalo finder respectively. Figure \ref{mass_function_fig} shows the halo mass functions predicted by \texttt{BlueTides} at $z=7.5$ for both FOF (0.2) and SO (200) halos. We find that the mass function for SO halos shows good agreement ($\lesssim 10\%$) with the analytical fit provided by \citep{2013ApJ...770...57B}. The mass function for FOF halos shows significant excess ($\gtrsim 50 \%$) at the very massive end ($M_H\gtrsim 3 \times 10^{11} M_{\odot}/h$). This occurs because these massive FOF groups are comprised of several collapsed objects (identified as distinct SO halos) "stiched" together. As an example to demonstrate the foregoing, the top left panel of Figure \ref{galaxy_snaps} shows the most massive FOF group at $z=8$. 

\section{Best fit HOD parameters and satellite concentrations}
Table \ref{HOD_parameter_tables} shows the best fit HOD parameters to the mean central and satellite occupations at redshifts $7.5-10$. The fitting is performed using the \texttt{scipy.optimize.curve_fit} package in python 2.7. The quoted errorbars are covariance errors, however these do not include errors due to cosmic variance.
\begin{table*}
\begin{tabular}{ccccccc}
\hline
Redshift & $\log_{10}M^*_{th}h M_{\odot}^{-1}$ & $\log_{10}M_{\mathrm{min}}h M_{\odot}^{-1}$ & $\sigma_{\log{M}}$ & $\log_{10}M_0 h M_{\odot}^{-1}$ & $\log_{10}M_1 h M_{\odot}^{-1}$ & $\alpha$ \\
\hline
7.5 & 8.0 & 10.456 & 0.20 & 10.26 & 11.38 +/- 0.02 & 0.99 +/- 0.03 \\
7.5 & 8.5 & 10.733 & 0.20 & 10.65 & 11.62 +/- 0.01 & 1.18 +/- 0.07 \\
7.5 & 9.0 & 11.018 & 0.21 & 10.93 & 11.86 +/- 0.01 & 1.49 +/- 0.11 \\
7.5 & 9.5 & 11.316 & 0.20 & 11.48 &  &  \\
7.5 & 10.0 & 11.619 & 0.18 &  &  &  \\
7.5 & 10.5 & 11.900 & 0.03 &  &  &  \\
\hline
8.0 & 8.0 & 10.466 & 0.21 & 10.28 & 11.35 +/- 0.01 & 0.87 +/- 0.03 \\
8.0 & 8.5 & 10.753 & 0.21 & 10.70 & 11.67 +/- 0.03 & 0.80 +/- 0.13 \\
8.0 & 9.0 & 11.050 & 0.22 & 11.12 & 11.92 +/- 0.22 & 1.03 +/- 0.46 \\
8.0 & 9.5 & 11.345 & 0.21 & 11.41 &  &  \\
8.0 & 10.0 & 11.664 & 0.16 &  &  &  \\
8.0 & 10.5 & 11.826 & 0.03 &  &  &  \\
\hline
9.0 & 8.0 & 10.458 & 0.21 & 10.38 & 11.30 +/- 0.04 & 0.65 +/- 0.11 \\
9.0 & 8.5 & 10.752 & 0.21 & 10.73 & 11.85 +/- 0.20 & 0.58 +/- 0.25 \\
9.0 & 9.0 & 11.042 & 0.21 & 11.19 &  &  \\
9.0 & 9.5 & 11.348 & 0.22 &  &  &  \\
9.0 & 10.0 & 11.562 & 0.03 &  &  &  \\
\hline
10.0 & 8.0 & 10.462 & 0.22 & 10.45 & 11.14 +/- 0.04 & 1.54 +/- 0.30 \\
10.0 & 8.5 & 10.756 & 0.22 & 10.84 & 11.22 +/- 0.03 & 2.39 +/- 0.81 \\
10.0 & 9.0 & 11.043 & 0.23 &  &  &  \\
10.0 & 9.5 & 11.398 & 0.16 &  &  &  \\
\hline
\end{tabular}

\caption{Best fit HOD parameter estimates for galaxy samples with various stellar mass thresholds in \texttt{BlueTides} at redshifts 7.5,8,9,10. The error bars are covariance errors. In the entries where the errors not quoted, they are $\lesssim 1\ \%$. For the blank columns, no good fit was able to be obtained to constrain the parameters.}
\label{HOD_parameter_tables}
\end{table*}

\begin{table}
\begin{tabular}{ccc}
$\log_{10}M^*_{th}h M_{\odot}^{-1}$ & $\log_{10}M_{H}h M_{\odot}^{-1}$ & $c_{\mathrm{sat}}$ \\
\hline
8.0 & 10.5 & $\rightarrow \infty$ (Power law) \\
8.0 & 11.0 & $\rightarrow \infty$ (Power law) \\
8.0 & 11.5 & 14.6 +/- 4.8 \\
8.0 & 12.0 & 10.2 +/- 4.1 \\
\hline
8.5 & 10.5 & $\rightarrow \infty$ (Power law) \\
8.5 & 11.0 & $\rightarrow \infty$ (Power law) \\
8.5 & 11.5 & 32.6 +/- 16.6 \\
8.5 & 12.0 & 14.0 +/- 5.4 \\
\hline
9.0 & 11.0 & $\rightarrow \infty$ (Power law) \\
9.0 & 11.5 & $\rightarrow \infty$ (Power law) \\
9.0 & 12.0 & 40.6 +/- 20.2 \\
\hline
9.5 & 11.5 & $\rightarrow \infty$ (Power law) \\
9.5 & 12.0 & $\rightarrow \infty$ (Power law) \\
\hline
\end{tabular}
\caption{Best fit satellite concentration parameters at $z=7.5$ for various halo mass bins and stellar mass thresholds. The bin widths of halo masses were $\delta{M_H hM_{\odot}^{-1}}=0.25$. If the rightmost column says $c_{\mathrm{sat}}\rightarrow \infty$, the corresponding profiles were effectively power-laws with exponent -3, also implying that the satellite concentrations are too high to be effectively constrained.}
\label{concentration_table}
\end{table}

\bibliography{references}
\end{document}